\documentstyle[12pt]{article}
\begin{document}
\topmargin 0pt
\oddsidemargin -3.5mm
\headheight 0pt
\topskip 0mm
\addtolength{\baselineskip}{0.20\baselineskip}
\vspace{1.0cm}
\begin{flushright}
SOGANG-HEP $197/95$
\end{flushright}
\vspace{1.0cm}
\begin{center}
{\Large \bf  On the Foundation of the Relativistic Dynamics} \\
{\Large \bf with the Tachyon }
\end{center}
\vspace{0.0cm}
\begin{center}
{Mu-In Park and Young-Jai Park}\\
\vspace{0.0cm}
{Department of Physics$^{*}$ and Basic Science Research Institute }\\
{Sogang University, C.P.O.Box 1142, Seoul 100-611, Korea}\\
\vspace{1.0cm}
{\large \bf ABSTRACT}
\end{center}

The theoretical foundation of the object moving faster than light in vacuum
({\it tachyon}) is still missing or incomplete. Here we present the classical
foundation of the relativistic dynamics including the tachyon. An anomalous
sign-factor extracted from the transformation of ${ \sqrt{1-u^{2}/c^{2} } }$
under the Lorentz transformation, which has been always missed in the usual
formulation of the tachyon, has a crucial role in the dynamics of the tachyon.
Due to this factor the mass of the
tachyon transforms in the unusual way although the energy and momentum,
which are defined as the conserved quantities in all uniformly moving systems,
transform in the usual way as in the case of the object moving slower than
light ({\it bradyon}). We show that this result can be also obtained from the
least action approach. On the other hand, we show that the ambiguities for the
description of the dynamics for the object moving with the velocity of light
({\it luxon}) can be consistently removed only by introducing a new dynamical
variable.
Furthermore, by using the fundamental definition of the momentum and energy
we show that the zero-point energy
for any kind of the objects, {\it i.e.}, the tachyon, bradyon, and luxon, which
has been known as the undetermined constant, should satisfy some constraints
for consistency, and we note that this is essentially another novel
relativistic
effect. Finally, we remark about the several unsolved problems.

\vspace{1.0cm}
\begin{flushleft}
May 1995 \\
$^{*}$ Electronic address: mipark, yjpark@physics.sogang.ac.kr \\
\end{flushleft}
\newpage
\begin{center}
{\large \bf I. INTRODUCTION } \\
\end{center}

According to the Einstein's special theory of relativity, the object moving
faster than light in vacuum [1] ({\it tachyon} after G. Feinberg [2]) can
violate the causality, {\it i.e.}, the relationship between the causes and
effects has
no absolute meaning for the observers moving uniformly with each others [3,4].
Moreover, when the formulas of the energy and momentum for the object moving
slower than light ({\it bradyon} [5]) are {\it directly } applied to the
tachyon, the energy and momentum of the tachyon become to be imaginary-valued
[4,6]
implying that the tachyon is believed to be never detected by our ordinary
bradyonic detectors if energy-momentum are also conserved in this
process [7]. These were main objections to the existence of
the tachyon. However, many years later, Bilaniuk-Deshpande-Sudarshan and
Feinberg (BDSF) had suggested that these two difficulties can be avoided [2]:
the first difficulty is avoided by introducing the {\it imaginary-valued rest
mass} of the
tachyon, which can not be ruled out in principle because the rest mass is not
directly measurable quantity unless the object can be brought to rest, and by
noting that there are no rest system for the tachyon according to the result of
the usual Lorentz transformation. The second difficulty is avoided by
introducing the so-called {\it reinterpretation principle} of `` the negative
energy state of the tachyon with reversed or backward time sequences '' to be
the positive energy state of the tachyon with ordinary or forward time
sequences. However, the theoretical foundation of the suggestions of BDSF
is still missing or incomplete although their suggestions were
discussed by various authors [8], and further development on the quantum theory
of the tachyon field theory [9] were followed. This is essentially due to the
fact that their suggestions rely heavily on the formulas of the bradyon.
In fact, the energy
and momentum formulas for the bradyon can not be directly applied to the
tachyon
because some of the logics in the derivations of the formulas can not be true
for the tachyon. Hence the separate derivations of the formulas for the
tachyon
from the first principles are required. Actually the incompleteness of the
usual
formulations was pointed out by several authors [10], but the complete
formulation is still absent.

In this paper we present the classical foundation of the relativistic dynamics
including the tachyon.
In Sec. II the principles of the relativistic mechanics
are reconsidered for the case that there are tachyons as well as the bradyons.
An anomalous sign-factor in the transformation of
${\sqrt{1-u^{2}/c^{2}}}$ under the Lorentz transformation, which is unimportant
for the bradyon and has been always missed in the usual formulation
of the tachyon, has now a crucial role in the dynamics of the tachyon. Due to
this novel
sign-factor it is found from the consideration of the collision process
that the proper-mass of the tachyon is not an absolute quantity in contrast to
that
of the bradyon but transforms as $\kappa'$ = sign$\{1-{\bf u \cdot v}/c^{2} \}
\kappa$. Furthermore, we
show that the tachyon's mass has a definite sign for a given speed in a
uniformly moving system, but has the sign changed depending on the tachyon's
velocity and the relative motion of the systems $S$ and $S'$.

On the other hand, by defining the momentum and energy as the conserved
quantities of the presumed forms in all uniformly moving systems, we
show that the zero-point energy $\epsilon_{a}$ for any kind of the
objects of the both tachyon and bradyon, which has been known as the
undetermined constant, should satisfy some constraints for consistency.
Furthermore,
especially for the case of $\epsilon_{a}= 0$ for all $a$ as in the usual
conventions, it is found that the energy and momentum,
for the tachyon as well as the bradyon, satisfy the usual four-vector
transformation such that the energy for the tachyon does not have the invariant
sign of the energy, but the sign can be changed depending on the object's
velocity and the relative motion of the two systems of coordinates $S$ and $S'$
in contrast to that of the bradyon. However, we note that this transformation
formulas can not be obtained from the usual formulation of the tachyon in
contrast
to the usual belief such that the original
motivation of the BDSF's reinterpretation principle, if it applies to our
Nature, can not be found in the
usual formulation but, only in our new formulation of the tachyon.

In Sec. III we develop the least action approach to the tachyon dynamics by
comparing the approach for the bradyon.
Although it is not clear whether this approach can be also applied
to the tachyon since the situation of $\Delta t = 0$ for the time
interval along the world-line can not be avoided for any tachyon such that
there are some ambiguities in deriving the equation of motion. However, we show
that this can be really also applied to the tachyon with the help of the
reparametrization invariance of the action, and the essential results for the
tachyon of the elementary approach of Sec. II can be also rederived in this
approach.

In Sec. IV we develop the least action approach for the object moving with the
velocity of light ({\it luxon}) [11]. We
emphasize that a new dynamical variable should be introduced as well as the
velocity variable in order to describe the luxon dynamics without any
ambiguity.
Their
physical meaning is uncovered by comparing the Einstein relation for the
photon.
It is shown that in this approach with the help of these new variables, the
usual energy-momentum relation for the luxon is proved without any ambiguity,
and the energy and momentum of the luxon also satisfy the usual transformation
like as the tachyon and bradyon, which has not been proved in the usual
classical
mechanical formulation.
Moreover, with the help of the energy and momentum transformation under the
Lorentz transformation, the zero-point energy for the luxon should satisfy the
same constraints as in the case of the tachyon and bradyon considered in
Sec. II, if we maintain the definition of the energy and momentum
as the conserved quantities for all the systems of coordinates and all the
possible process of the collision like as the case of the tachyon and bradyon.
Especially, for the case of the photon it's zero-point energy should be zero
due
to the fact that it is clearly {\it the strictly neutral particles}.

Sec. V is devoted to summary and concluding remarks especially on
the problems of causality violation, the \v{C}erenkov effect in
the vacuum, and the quantization.\\

\begin{center}
{\large \bf II. THE PRINCIPLES OF RELATIVISTIC MECHANICS INCLUDING THE TACHYON
}
\\
\end{center}

The space and time of a system of coordinates, in which the equations of
Newtonian mechanics hold well, and which will be denoted by stationary system
$S$
hereafter, are essentially defined by the measuring rods and clocks at rest
relatively to this system {\it without references} to the motions and
properties
of the measured object. Moreover, the transformation law of the space and
time from a system $S$ to another system $S'$ in uniform motion
of translation relatively to the former with the relative speed
to be smaller than that of light, is uniquely determined by postulating
a) the two basic hypothesis, {\it i.e.}, the principle of relativity and
constancy of the velocity of light with b) some plausible assumptions on the
structure of the space and time, {\it i.e.}, the homogeneity of space and time,
the isotropy of space and the validity of Euclidean geometry on this space and
time, and c) tacit assumption that there exists
at least one stationary system relatively to the observers for any motion of
them.

In the usual formulation of the relativistic mechanics, if the momentum of the
point objects
\begin{eqnarray}
{\bf p}=m({\bf u})~{\bf u}
\end{eqnarray}
is essentially defined to be the conserved quantity in total for
any stationary system $S$, the mass function $m({\bf u})$ is
found to be
\begin{eqnarray}
m({\bf u})=\frac{m_{0}}{\sqrt{ 1- u^{2}/c^{2} }}.
\end{eqnarray}
Moreover, the quantity of the
energy having the form
\begin{eqnarray}
E=m({\bf u})~c^{2}
\end{eqnarray}
is proved to be totally conserved [12], and as a result
the principle of energy conservation is not independent on the principle of
momentum conservation [13]. However, some of the usual methods of derivations
of
these formulas are valid
only for the bradyon case essentially due to the
non-existence of the rest system for the tachyon [13]. Therefore, we must
carefully select the methods of
derivation in order to include, from the start, the tachyon as
well in the process
of interaction of objects.  Fortunately, there exists already an appropriate
formulation for this purpose although the power of this formulation has
been never well noticed [14]. Rather, this latter formulation has been believed
to be worse than the former in the aspect of the economics of the logic because
the
latter
formulation assumes one more condition, which is actually equivalent to the
principle of energy conservation, on the mass function $m({\bf u})$ in order to
derive all the same results of the former formulation [15]. But, now, by using
the latter formulation, we can discuss the mechanics of the tachyon from the
start without any problem at the cost of the economy of the logic.\\

{\bf A}. {\bf The definition of the mass and the results drawn from it}\\

The theory of the mechanics can only be constructed by defining the mass of
inertia
of the point objects. In order to define the mass, let $S$ and $S'$ be two
systems of coordinates with the relative velocity ${\bf v}$, and consider a
system of point objects composed of the tachyon as well as the bradyon
in both $S$ and $S'$. We fundamentally define the mass of inertia $m$ of
each object in the system $S$ such that both
\begin{eqnarray}
\sum_{a}m_{a}
\end{eqnarray}
and
\begin{eqnarray}
\sum_{a} m_{a} {\bf u}_{a}
\end{eqnarray}
are conserved through all the possible processes of collision and in all the
inertial
systems of coordinates. Then, by representing ${\bf u}'_{a}$ as the velocity of
$a$-th object in $S'$, the mass of inertia of each object in $S'$
is defined such that both
\begin{eqnarray}
\sum_{a}m_{a}'
\end{eqnarray}
and
\begin{eqnarray}
\sum_{a} m_{a}' {\bf u}_{a}'
\end{eqnarray}
are also conserved through all the possible processes of collision.

In order to find the explicit formula of the mass we first note that
\begin{eqnarray}
\nonumber
\sqrt{1-{u'}_{a} ^{2}/ c^{2} }&=&\frac{\sqrt{1-u_{a} ^{2} / c^{2}}
{}~\sqrt{1-v^{2}/c^{2}}  }
                   {|1-{\bf u}_{a} \cdot {\bf v} /c^{2} |} \\\\ \nonumber
            &=&\frac{ \sqrt{ 1-u_{a} ^{2} / c^{2} }~ \sqrt{ 1-v^{2}/c^{2} }}
{ \{1-{\bf u}_{a} \cdot {\bf v} /c^{2} \}
                 ~\mbox{sign}\{1-{\bf u}_{a} \cdot {\bf v} /c^{2} \} }
\end{eqnarray}
from the well known usual velocity transformation formula
\begin{eqnarray}
{\bf u}'_{a}=~\frac{ \sqrt{ 1-v^{2}/c^{2} }~{\bf u}_{a}
                     ~+~ ( {\bf u}_{a} \cdot {\bf v} )~{\bf v}
{}~[1-\sqrt{1-v^{2}/c^{2}}]/v^{2} - {\bf v} }
{ 1-{\bf u}_{a} \cdot {\bf v} /c^{2} },
\end{eqnarray}
where we assume that if an object is observed in one system, then this object
also
should be observed in another system, which is uniformly moving relative to the
former such that Eqs. (8) and (9) are always meaningful.
Moreover, we only consider here the case of $v <c$, {\it i.e.}, bradyonic
motion of the system of coordinates since it is remainly doubtful whether the
physically acceptable Lorentz transformation for the case of $v~>~c$ exists or
not [16]. Most importantly, the sign-factor of
Eq. (8), which has no effect for the usual treatment of the bradyon,
is now the novel effect of the tachyon. Then, using an identity from Eq. (8)
\begin{eqnarray*}
1=\gamma
 ~\frac{\sqrt{1-{u'}_{a} ^{2}/ c^{2} } }
            { \sqrt{1-u_{a} ^{2} / c^{2}} }
{}~\left\{
1- \frac{  {\bf u}_{a} \cdot {\bf v}  } { c^{2} }   \right\}
{}~\mbox{sign}
\left\{ 1-\frac{ {\bf u}_{a} \cdot {\bf v}  } { c^{2} }      \right\},
\end{eqnarray*}
where $\gamma~=~ 1/\sqrt{1-v^{2}/c^{2}}$, and Eq. (9), Eqs. (6) and (7) can be
rewritten as follows
\begin{eqnarray}
\nonumber
\sum_{a} m'_{a}&=& \sum _{a}
m'_{a}~\left[\gamma ~\frac{\sqrt{1-{u'}_{a} ^{2}/ c^{2} } }
            { \sqrt{1-u_{a} ^{2} / c^{2}} }
{}~\left\{
1- \frac{  {\bf u}_{a} \cdot {\bf v}  } { c^{2} }   \right\}
{}~\mbox{sign}
\left\{ 1-\frac{ {\bf u}_{a} \cdot {\bf v}  } { c^{2} }      \right\}
\right]              \\\\ \nonumber
&=&\gamma
\left[
\sum_{a} m_{a} - \frac{  {\bf v} } {c^{2}} \cdot \sum_{a} (m_{a} {\bf u}_{a})
\right] ,  \\  \nonumber  \\ \nonumber
\sum_{a} (m'_{a} {\bf u}'_{a} )&=&\sum_{a}
m'_{a}~\left[
 \frac{\sqrt{1-{u'}_{a} ^{2}/ c^{2} } }
     { \sqrt{1-u_{a} ^{2} / c^{2}}  }
\left\{
{\bf u}_{a}+ ( \gamma-1 )
\frac{( {\bf u}_{a} \cdot {\bf v} ) ~{\bf v} } { v^{2} }
- \gamma {\bf v}
\right\} \right.  \\ \nonumber  \\
{}~~~~~~&&\left.
\times \mbox{sign}
\left\{ 1- \frac{  {\bf u}_{a} \cdot {\bf v}  } { c^{2} }       \right\}
\right]   \\ \nonumber \\ \nonumber
&=& \left[ 1 + ( \gamma-1) \frac{  ~{\bf v} }{v^{2}}  {\bf v}~\cdot ~ \right]
\sum_{a}(m_{a}{\bf u}_{a})
-\gamma{\bf v} \sum_{a} m_{a},  \nonumber
\end{eqnarray}
where
\begin{eqnarray}
m_{a}~=m'_{a}~
 \frac{\sqrt{1-{u'}_{a} ^{2}/ c^{2} } }
     { \sqrt{1-u_{a} ^{2} / c^{2}}  }
{}~\mbox{sign}
\left\{ 1-\frac{ {\bf u}_{a} \cdot {\bf v}  } { c^{2} }     \right\}.
\end{eqnarray}

Note that the transformation laws (10) and (11) are uniquely determined by
applying our definition of the mass, {\it i.e.}, if
$\sum_{a} m_{a}$ and $\sum_{a}m_{a} {\bf u}_{a}$ are conserved in $S$, this
should
be also true for $\sum_{a}m'_{a}$ and $\sum_{a} m'_{a}{\bf u}'_{a}$ in $S'$.
Hence,
the mass transformation law (12) is unique. Without the sign factor in Eq.
(12),
Eqs. (10) and (11) show that the conservation of
$\sum_{a}m_{a}$ and $\sum_{a} m_{a} {\bf u}_{a}$ does not always imply the
conservation of $\sum_{a}m'_{a}$ and $\sum_{a}m'_{a}{\bf u}'_{a}$.

Now, it remains to obtain the mass function $m$ satisfying the mass
transformation (12) as a function of dynamical variables. To this end, we first
consider the case of the bradyon, {\it i.e.}, $u^{2}_{a}~<~c^{2}$ or
equivalently $u'^{2}_{a}~<~c^{2}$. In this case, the mass transformation (12)
is reduced to be
\begin{eqnarray}
m_{a}~=m'_{a}~
 \frac{\sqrt{1-{u'}_{a} ^{2}/ c^{2} } }
     { \sqrt{1-u_{a} ^{2} / c^{2}}  },
\end{eqnarray}
or
\begin{eqnarray}
m_{a}~\sqrt{1-u_{a} ^{2} / c^{2}} ~=m'_{a}~
 \sqrt{1-{u'}_{a} ^{2}/ c^{2} } =\mbox{frame independent constant.}
\end{eqnarray}
By considering especially the object rest system with $u_{a}=0$, we
can see that the frame independent constant, which should be an inherent
property of the object, is nothing but the object's mass when it is at rest,
{\it i.e.,} rest mass $m_{0}$. Then, the mass for the non-zero velocity objects
can be obtained as
\begin{eqnarray}
m_{a}({\bf u}_{a})~=~\frac{ m_{0~a} }{ \sqrt{1-{u}_{a} ^{2}/ c^{2} }  }.
\end{eqnarray}
Note that this does not depend on the direction of the objects' velocity, but
only it's magnitude. Actually this property can be easily understood using the
isotropy of the space. Suppose that there is a rest object in a system of
coordinates $S$. Then, if there is the mass relation between the rest object
and
the object moving relative to the rest one for the same kind objects, the
transformation function should not depend on the direction of the velocity of
the moving object since all the directions relative to the rest object are
equivalent due to the isotropy of the space (Fig.1).

On the other hand, the situation is very different for the case of tachyon,
{\it i.e.}, $u_{a}^{2}~>~c^{2}$ or equivalently ${u'}_{a}^{2}~>~c^{2}$. In
this case {\it the anomalous sign-factor} in the mass transformation (12)
becomes
important. Due to this novel sign-factor, the quantity
$m_{a}\sqrt{1-u_{a} ^{2} / c^{2}}$ is no more the frame independent one. Rather
the true frame independent quantity is the squared quantity of this, {\it
i.e.,}
\begin{eqnarray}
m_{a}^{2}~\{u_{a} ^{2} / c^{2}-1\} ~={m'}_{a}^{2}~
 \{{u'}_{a} ^{2}/ c^{2}-1 \} =\mbox{frame independent constant.}
\end{eqnarray}
Now, by considering the system of coordinates when the object's velocity is
$u_{a}^{2}=2 c^{2}$, we can see that the frame independent constant, which
should be also an inherent property of the tachyonic object, is nothing but
the square of the objects' mass at $u_{a}^{2}=2 c^{2}$.
Then the square of the mass for the general velocity of the tachyon is
determined as
\begin{eqnarray}
m_{a}^{2} ({\bf u}_{a})~=~\frac{ \mbox{square~ of~mass~ at}~u_{a}^{2}=2 c^{2}
}
{ u_{a}^{2} / c^{2}-1   }.
\end{eqnarray}
But this formula does not determine the sign of the mass, {\it i.e.}, when the
mass function is expressed, without destroying the generality as follows
\begin{eqnarray}
m_{a}~=~\frac{ \kappa_{a} }{ \sqrt{{u}_{a}^{2}/ c^{2}-1 }  },
\end{eqnarray}
the sign of $\kappa_{a}$, which may depend on it's motion, is not determined
but only $\kappa_{a}^{2}$ is found to be frame independent.

Before we discuss
the determination of the sign-factor, it seems appropriate to show that the
mass
function $m_{a}$ and hence $\kappa_{a}$ may depend on the direction of the
velocity in contrast to the case of the bradyon. This can be easily understood
by
the fact that any tachyon's mass is determined by comparing with the mass of
the
object with the same kinds having the velocity $u_{a}^{2}=2 c^{2}$ and a
definite direction. Hence, all the directions of the velocity of the general
tachyon are not equivalent due to the asymmetry of the situation. So if there
is any relation between the mass of the general tachyon with that of the
tachyon
of $u_{a}^{2}=2 c^{2}$, the function can be depend on the velocity of the
tachyon relative to the tachyon of $u_{a}^{2}=2 c^{2}$ in general (Fig. 2).

Now we discuss on the sign problem of the mass. To this end we first note that
$\kappa_{a}$, which is called the proper mass of the tachyon, can be expressed
as
\begin{eqnarray}
\kappa_{a}=|\kappa_{a}|\cdot (\mbox{sign-function}),
\end{eqnarray}
where the magnitude of $\kappa,~|\kappa_{a}|$, does not depend on the direction
of the tachyon's velocity since $\kappa^{2}_{a}$ is the frame independent
constant, and hence only the ``sign-function'' may depend on the
direction of the velocity as guaranteed by the asymmetry of the situation of
the
mass determination method. As for the sign-function, we first note that the
tachyon's mass and hence the sign-function do not depend on the direction of
the tachyon for a given speed. This can be understood by considering the
collision process of the tachyon with the bradyon. If the mass of the tachyon
depends on the direction of the velocity for a given speed, it should also be
changed for the different orientations of the coordinates (Fig. 3). But, this
is
not allowed in order that the total energy and momentum of Eq.(4) and (5) are
conserved in any orientation of the coordinates because the bradyon's mass is
independent on the direction of the velocity. Hence, the tachyon's mass has a
definite sign for a given speed in a uniformly moving system. However, the sign
itself can not be determined {\it a priori} but only by actual measurement
for a given situation. For the method of
determination of the mass, we can imagine that the sign is uniquely determined
by considering the collision-process between the tachyons having the given
speed, with the bradyon operationally since the bradyon has a definite sign of
the mass for any motions. However, for the different velocity the sign of the
mass can be changed depending on the direction and the magnitude of the
tachyon's velocity relative to the relative motion of the system of coordinates
according to Eq. (12). In the expression of the tachyon's proper mass
$\kappa_{a}$, the Eq. (12) implies
\begin{eqnarray}
{\kappa}_{a}~=~{\kappa}'_{a}~\mbox{sign}
\left\{ 1-\frac{ {\bf u}_{a} \cdot {\bf v}  } { c^{2} } \right\}
\end{eqnarray}
since the mass in the system $S'$ also should be expressed as
\begin{eqnarray}
m_{a}'~=~\frac{ \kappa_{a}' }{ \sqrt{{u'}_{a}^{2}/ c^{2}-1 }  }
\end{eqnarray}
due to the principle of relativity. Hence the proper mass of the tachyon is not
an absolute quantity in contrast to that of the bradyon. This is {\it the novel
feature}
of the tachyon. Furthermore, we note that although the sign-factor of Eq. (20)
is ill-defined for the case of $1-{\bf u}\cdot {\bf v}/c^{2} =0$, {\it i.e.,}
$u'_{a}=\infty$ from Eq. (8) or (9) such that the sign of $\kappa'_{a}$ is
ill-defined in this case,
the directly measurable quantity $m'_{a}$ is well-defined according to Eq.
(21):
$m'_{a}$ becomes zero in this case for any finite $\kappa '_{a}$, which is
expected to be the only physically realizable case.\\

{\bf B}. {\bf Energy-momentum transformation law and it's implications }\\

We define momentum as the quantity
\begin{eqnarray}
{\bf p}_{a}=m_{a}({\bf u}_{a})~ {\bf u}_{a},
\end{eqnarray}
which is conserved through all possible process of collision and in all systems
of
coordinates for the $a$'th object.

Moreover, we fundamentally define the (kinetic) energy $E_{a}$ of the object
such that
the change of the mass of the object between one state of motion $A$ and
another state of motion $B$ is completely spent to
change the energy of the object, {\it i.e.},
\begin{eqnarray}
m_{a}(B)~c^{2} -m_{a}(A)~c^{2} =E_{a(B)}-E_{a(A)}.
\end{eqnarray}
Then, the energy has the expression as
\begin{eqnarray}
E_{a}=m_{a}({\bf u}_{a})~c^{2} +\epsilon_{a},
\end{eqnarray}
where $\epsilon_{a}$ is the integration constant, which is inherent to the
object and hence may be different for the different species of object
generally, and is related to the choice of the origin of the energy such that
it
may be expected that they have arbitrary values [4]. However, by defining
the energy also as the conserved quantity for all the possible processes of
collision and all the systems of coordinates, we can obtain some constraints
about the value of $\epsilon _{a}$. To this end, let us first note
that Eq. (11) can be rewritten as
\begin{eqnarray}
\sum_{a}~(m'_{a} {\bf u}'_{a})&=&
\left[ 1 + ( \gamma-1) \frac{  ~{\bf v} }{v^{2}}  {\bf v}~\cdot ~ \right]
        \sum_{a}(m_{a}{\bf u}_{a} )
             \nonumber \\\\  \nonumber
   && -~\gamma \frac{{\bf v}}{c^{2}}  \sum_{a}~ (m_{a}c^{2}
   +\epsilon_{a})  +
    ~\gamma \frac{{\bf v}}{c^{2}}  \sum_{a} \epsilon_{a}
\end{eqnarray}
or
\begin{eqnarray}
\sum_{a}{\bf p}'_{a}&=& \left[ 1 + ( \gamma-1) \frac{  ~{\bf v} }{v^{2}}
{\bf v}~\cdot ~ \right] \sum_{a}{\bf p}_{a}
                     \nonumber \\\\ \nonumber
   && -~\gamma \frac{{\bf v}}{c^{2}}  \sum_{a} E_{a}  +
     ~\gamma \frac{{\bf v}}{c^{2}}  \sum_{a} \epsilon_{a}.
\end{eqnarray}
Now, in order that through the process of collision the total momentum is
conserved in $S'$, we must also have the conservation of $\sum_{a}
\epsilon_{a}$, {\it i.e.,}
\begin{eqnarray}
\sum_{a_{i}=1}^{M} \epsilon_{a_{i}}
=\sum_{a_{f}=1}^{N} \epsilon_{a_{f}}
\end{eqnarray}
for the $In$ state $i$ and $Out$ state $f$, as well as the total energy and
momentum in $S$ (Fig. 4). Here it is easy to see, by considering the
pair-creation processes, that $\epsilon_{a}=0$ for the
$strictly~ neutral~ particles$, which are defined as the object having
the property of
`` particle = antiparticle '' [17], but $\epsilon _{a~(\mbox{\scriptsize
particle})}
= -\epsilon _{a~ (\mbox{\scriptsize antiparticle})}$ in the other cases. In the
latter
cases, the value of $\epsilon_{a}$ itself is remainly undetermined.
Furthermore, note that this is essentially the relativistic effect, which has
not
been known before: the last term in Eq. (26), which is crucial to this effect,
does not appear in the non-relativistic limit of $v \rightarrow 0$.

Independently with this argument,
we consider, now, the transformation law of Eq. (10). By noting that Eq. (10)
can be reexpressed as
\begin{eqnarray}
\sum_{a}(m'_{a}c^{2}+\epsilon '_{a})-\sum_{a} \epsilon '_{a} =\gamma
\left[ \sum_{a}~(m_{a}c^{2} +\epsilon_{a})-\sum_{a} \epsilon_{a}
                     -{\bf v} \cdot \sum_{a}~(m_{a}{\bf u}_{a})  \right],
\end{eqnarray}
we obtain the following relation
\begin{eqnarray}
\sum_{a}E'_{a}-\sum_{a} \epsilon '_{a}=\gamma \left[ \sum _{a} E_{a}
        -{\bf v} \cdot \sum_{a}{\bf p} -\sum_{a} \epsilon_{a} \right],
\end{eqnarray}
where
\begin{eqnarray}
E'_{a}=m'_{a}c^{2} +\epsilon'_{a}.
\end{eqnarray}
Hence the total energy is also conserved in $S'$ through the process of
collision, when the total energy and momentum are conserved in $S$
only if the integration constant is also conserved in $S'$ since we already
know it's conservation in $S$ form Eq. (27). Due to this conservation of
$\epsilon '_{a}$ in $S'$ the constraints on it's value described in the below
Eq. (27)
are universally valid in all systems of coordinates.

Furthermore, note that these results for the property of $\epsilon_{a}$ is
valid for both the tachyon and the bradyon. It will be shown later in
Section IV that these are also true for the case of the luxon.

When expressed with the total momentum and energy, Eqs. (26) and (29) with
$\epsilon_{a}\equiv 0$ as in the usual treatment, now become
\begin{eqnarray}
{\bf p}'&=&
\left[ 1 + ( \gamma-1) \frac{  ~{\bf v} }{v^{2}}  {\bf v}~\cdot ~ \right]
{\bf p}
      -\gamma \frac{{\bf v}}{c^{2}} E,  \\ \nonumber
      \\
E'&=& \gamma~[ E-{\bf p} \cdot {\bf v} ],
\end{eqnarray}
where
\begin{eqnarray}
{\bf p}=\sum_{a}m_{a} {\bf u}_{a}&,&~ E=\sum_{a} m_{a} c^{2}, \\ \nonumber
{\bf p}'=\sum_{a}m'_{a} {\bf u}'_{a}&,&~ E'=\sum_{a} m'_{a} c^{2},
\end{eqnarray}
and the single point object also satisfies the same transformation law.
The transformation law (32) shows that the sign of the (total)
energy may be changed when the (center of mass) velocity ${\bf u}$ is faster
than that of light because
\begin{eqnarray}
E'= \gamma~ E
      \left\{1-\frac{{\bf u} \cdot {\bf v}}{c^{2}} \right\}
\end{eqnarray}
for the (center of mass) single object such that ${\bf p}=(E/c^{2}){\bf u}$.

Although this has been known as the novel feature of the usual tachyon
theory, but this is actually not the case, {\it i.e.}, this phenomena can not
be obtained in the usual formulation. To see this we note that in the usual
formulation [2,8,9-11] the momentum and energy of the point objects are assumed
to be
\begin{eqnarray}
{\bf p}&=&\frac{\mu ~{\bf u}}{\sqrt{u^{2}/c^{2}-1}}, \\
E&=&\frac{\mu ~ c^{2}}{\sqrt{u^{2}/c^{2}-1}},
\end{eqnarray}
where $\mu$ is constant, and is the same for all system of coordinates.
The latter point is essentially due to the fact that the formulas (35) and (36)
are obtained from the usual formulas (1)-(3) of the bradyon by
substituting simply $m \rightarrow i \mu$ such that ${\bf p}$ and $E$ are
real-valued for the velocity faster than that of light. With the help of the
formula
(8) we can easily see that
\begin{eqnarray}
{\bf p}'&=&
\left\{ \left[ 1 + ( \gamma-1) \frac{  ~{\bf v} }{v^{2}}  {\bf v}~\cdot ~
\right]
       {\bf p}
     -\gamma~ \frac{ {\bf v} }{ c^{2} } ~E \right\}
     ~\mbox{sign} \left\{1-\frac{{\bf u} \cdot {\bf v}}{c^{2}} \right\}
     \nonumber
     ,   \\ \nonumber \\
E'&=& \gamma~( E-{\bf p} \cdot {\bf v} )~
     \mbox{sign} \left\{1-\frac{{\bf u} \cdot {\bf v}}{c^{2}} \right\}.
\end{eqnarray}
Hence the energy for the single object case becomes
\begin{eqnarray}
E'= \gamma~ E~
      \left|1-\frac{{\bf u} \cdot {\bf v}}{c^{2}} \right|
\end{eqnarray}
showing that the sign of the energy does not change for any motion of the
tachyon. Therefore the usual belief that the so called ${\it reinterpretation~
principle}$ for `` the state of tachyon of negative energy and reversed time
sequence '' of BDSF is needed to avoid the problem of
the negative energy simultaneously with the causality problem is
misunderstanding due to their wrong formulas of the usual formulation for the
tachyon. Hence the original motivation for the BDSF's reinterpretation
principle, if it applies to our Nature, can not be found in the usual
formulation but only in our new
formulation of the tachyon.  Furthermore, as it is clear from
the formula (37), total energy and momentum may be non-conserved in
one system of coordinates although conserved in the other systems within the
usual formulation [18]. It
follows from this result that the tachyon can not be represented by simple
replacement $m \rightarrow i \mu$ in contrast to the usual belief.

In addition we would like to remark that although there is the anomalous
sign-factor in the mass transformation formulas (12) or (21),
$E^{2}-c^{2}p^{2}$
is Lorentz invariant, and $(E, c{\bf p} )$ forms a $four-vector$ for the
tachyon
as well as the bradyon. \\

\begin{center}
{\large \bf III. LEAST ACTION APPROACH TO THE TACHYON DYNAMICS} \\
\end{center}

In the previous section we have considered the tachyon dynamics in the most
elementary way. Now, we reconsider the topics of the previous section in a
more mathematical framework, {\it i.e.}, the least action principle approach,
and the results of the previous section will be rederived in this formulation.
It is well known that the relativistic dynamics of point bradyon can be
elegantly described by the least action principle. Moreover, remarkably the
formulation of the free bradyon
can be uniquely described by considering some Lorentz invariant quantities and
presumed properties of the free objects, {\it i.e.}, the only dynamical
variables should be the velocity ${\bf u}$ of these objects although the
interactions with other fields of force are not uniquely defined unless more
informations or restrictions are not given [19].

However, as it will be clear later, the applicability of the approach for the
tachyon and luxon is not trivial matter. These problem will be treated in this
and next section, where it will be shown that the least action principle is
also applicable
to the case of the tachyon and luxon.\\

{\bf A}. {\bf The world-parameter}\\

For the usual bradyonic object there is an invariant parameter $\tau$, which is
called the proper time of this object, such that
\begin{eqnarray}
c ~d \tau &\equiv & c~ dt ~\sqrt{1-u^{2}/c^{2}} = c~
dt'~\sqrt{1-{u'}^{2}/c^{2}}
   \nonumber \\
                  & = &c~ d \tau '
\end{eqnarray}
with the help of the usual velocity transformation of the bradyon and the
parameter
very useful to parameterize the world-trajectories of the relativistic point
objects with the Lorentz invariant fashion. So we can expect that it would be
also useful to introduce similar parameter for the world-trajectory of the
tachyon in the formulation.

To this end, let us introduce the world-parameter ${\it w}$ describing
the world-trajectories of the relativistic point objects including the tachyons
as well as the bradyons unifiedly by the infinitesimal increment
\begin{eqnarray}
c ~d{\it w} \equiv   dt \sqrt{d x^{\mu} d x_{\mu}/(dt)^{2} }
             = c ~dt \sqrt{1-u^{2}/c^{2}}
\end{eqnarray}
for the infinitesimal interval in the world-trajectory of the object. Here we
use $x^{\mu} \equiv (ct, {\bf x}),~x_{\mu} =\eta_{\mu \nu}x^{\nu}
= (ct,- {\bf x})$, and $\eta_{\mu \nu}=diag(+1,-1,-1,-1)$.
By the general
velocity transformation law (8) and the usual Lorentz transformation, which are
also valid for tachyon as well as the bradyon, the transformation law of $c~dw$
becomes
\begin{eqnarray}
c ~d{\it w'} & = &c~ dt' \sqrt{ 1-{u'}^{2}/c^{2} } \nonumber \\
     &=&c ~\gamma \left\{ dt -\frac{ {\bf v}\cdot
     d{\bf x} }{c^{2}} \right\}
     \frac{ \sqrt{1-u^{2}/c^{2}}~   }
                          { \gamma |1-{\bf u} \cdot {\bf v} /c^{2} | }
\nonumber
                          \\\\
     \nonumber
     &=&c ~dt \sqrt{1-u^{2}/c^{2} } ~\frac{ \{ 1 -{\bf u}\cdot {\bf v}/{c^{2}}
     \} }
                               { |1-{\bf u} \cdot {\bf v} /c^{2} | }  \\
                               \nonumber \\  \nonumber
     &=&\mbox{sign} \{ 1 -\frac{{\bf u}\cdot {\bf v}}{c^{2}} \} c~ d{\it w}  .
\end{eqnarray}
Therefore, the world-parameter transforms as
\begin{eqnarray}
 {\it w'}=\mbox{sign} \{ 1 -\frac{{\bf u}\cdot {\bf v}}{c^{2}} \}{\it w}
\end{eqnarray}
by choosing appropriate origins of ${\it w}$ and ${\it w'}$ using the
homogeneity of the time $t$. Note that ${\it w}$ is not Lorentz invariant any
more, but a pseudo-Lorentz invariant, {\it i.e.}, Lorentz invariant up to a
sign-factor [20], while ${\it w}^{2}$ is the Lorentz invariant as in the usual
case. Especially, for bradyonic case the world-parameter ${\it w}$ itself
becomes to be Lorentz
invariant for any uniform motion of the system, and is reduced to the
usual proper time $\tau$, {\it i.e.}, the time measured by a clock which being
rest at the objects. Although this interpretation is not possible for tachyon
case, but we can say that ${\it w}$ has the same value of the time (interval)
of
the tachyon measured by a clock (of course, being made of bradyons) whose
velocity is $\sqrt{2} c$ relative to the tachyon. Moreover, in this case
if we introduce the real valued world-parameter $\lambda$ by
$\lambda \equiv -i {\it w}$ such that
\begin{eqnarray}
c~ d\lambda =c~ dt ~\sqrt{u^{2}/c^{2}-1},
\end{eqnarray}
then $\lambda$ transforms as
\begin{eqnarray}
\lambda' =\mbox{sign} \{ 1 -\frac{{\bf u}\cdot {\bf v}}{c^{2}} \} \lambda ,
\end{eqnarray}
where the anomalous sign factor can not be neglected. Here we note that
the real parameters $\tau$ and $\lambda$ can be used to parameterize the length
of the world trajectories of bradyon and tachyon, respectively.

Before ending this subsection it seems appropriate to comment on our choice of
the world-parameter as Eq. (40).
In the usual formulation of the relativistic point bradyon, the two definitions
\begin{eqnarray}
c~d \tau \equiv dt \sqrt{(d x^{\mu} dx_{\mu})/(dt)^{2}} =c~dt
\sqrt{1-u^{2}/c^{2}}
\end{eqnarray}
and
\begin{eqnarray}
c~d \hat{\tau} & \equiv & \sqrt{d x^{\mu} dx_{\mu}}
              = |dt| \sqrt{(d x^{\mu} dx_{\mu})/(dt)^{2}} \nonumber \\
         &=&c~|dt| \sqrt{1-u^{2}/c^{2}}
\end{eqnarray}
are interchangeably used. This is essentially due to the fact that, for the
case of the
bradyon the
sign of $dt$ and hence the order of events are Lorentz invariant. Due to this
fact, both $\tau$ and $\hat{\tau}$ equally describe the trajectory
invariantly. Note that the time direction of the world-parameter can be chosen
to be the
same as that of $t$ for $\tau$, or always positive for both the positive and
negative direction of $t$ for $\hat{\tau}$, and moreover this assignment of the
sign
is Lorentz invariant. Hence, in this case the two parameters have the
equivalent
role. Mathematically, this fact can be expressed by the fact that both $\tau$
and $\hat{\tau}$ are Lorentz invariant [19].

However, this situation is drastically changed for the case of the tachyon. In
this case $w$
in Eq. (40) and other definition similar to Eq. (46) as
\begin{eqnarray}
c~d \hat{w} & \equiv & \sqrt{d x^{\mu} dx_{\mu}}
      = |dt| \sqrt{(d x^{\mu} dx_{\mu})/(dt)^{2}} \nonumber \\
         &=&c~|dt| \sqrt{1-u^{2}/c^{2}}
\end{eqnarray}
have different role each other, and so can not be used interchangeably. Since
the
sign of
$dt$ for the tachyon is not Lorentz invariant, $w$ is not Lorentz invariant
although $\hat{w}$ is. However, essentially because of this fact, $\hat{w}$ can
not be used as the world-parameter for the tachyon. Note that since $\hat{w}$
is blind for the
time reversion, which is always possible for the tachyon, $\hat{w}$
can not describe the time reversion phenomena of the tachyon, but $w$ is not
the
case.
So we can conclude that the definitions (40) and (45) are more fundamental than
those of (46) and (47). This
is the reason why we does not introduce the Lorentz invariant parameter
$\hat{w}$ as the world-parameter but introduce the parameter $w$, which is not
Lorentz invariant, as in Eq. (40). This situation is comparative to that for
the
classical theory of the pair-creation or annihilation of a particle and it's
anti-particle of St\"{u}eckelberg and Feynman [21], where $\hat{\tau}$
is more favored than $\tau$.
\\

{\bf B}. {\bf The world-velocity and acceleration}\\

Now, using the world-parameter ${\it w}$ we can define the world-velocity
${\cal U}^{\mu}$ describing the tachyon as well as the bradyon unifiedly,
as follows
\begin{eqnarray}
{\cal U}^{\mu} \equiv \frac{d x^{\mu}}{d \it{w}}.
\end{eqnarray}

Then, due to the four-vector property of the quantity $d x^{\mu}$ and the
pseudo-scalar property of $d \it{w}$, the world-velocity $\cal{U}^{\mu}$
becomes the quantity of the pseudo four-vector, and it's components are given
by
\begin{eqnarray}
\cal{U}^{\mu} &=&\left( \frac{c~ dt}{d \it{w}},
              \frac{d {\bf x}}{d \it{w}} \right) \nonumber \\
&=&\left( \frac{c}{\sqrt{1-u^{2}/c^{2}}},\frac{{\bf u}}{\sqrt{1-u^{2}/c^{2}}}
 \right),
\end{eqnarray}
where ${\bf u}\equiv d {\bf x}/dt$.
This transforms under the Lorentz transformation as
\begin{eqnarray}
\nonumber \\
\cal{U'}^{\mu} &=&\left( \frac{c~ dt'}{d \omega '},
              \frac{d {\bf x}'}{d \omega '} \right) \nonumber \\ \nonumber \\
              \nonumber
    &=&\frac{1}{\mbox{sign} \{1 -{\bf u}
       \cdot {\bf v}/{c^{2}} \}}~ {\Lambda ^{\mu}}_{\nu} ~\left( \frac{c~ dt}
       {d \omega },
              \frac{d {\bf x}}{d \omega } \right) \nonumber \\ \nonumber \\
    &=&\frac{1}{\mbox{sign} \{1 -{\bf u}
       \cdot {\bf v}/{c^{2}} \}}~ {\Lambda ^{\mu}}_{\nu}
    ~{\cal U}^{\nu},     \\ \nonumber
\end{eqnarray}
where ${\Lambda^{\mu}}_{\nu}$ is the general Lorentz transformation matrix, for
the velocity ${\bf v}$ of the system $S'$ relative to the system $S$,  without
rotation as
\begin{eqnarray}
d{x'}^{\mu}={\Lambda^{\mu}}_{\nu} d x^{\nu},
\end{eqnarray}
with components
\begin{eqnarray}
{\Lambda^{0}}_{0}&=&1/\sqrt{1-v^{2}/c^{2}}, \nonumber \\
{\Lambda^{i}}_{0}&=&{\Lambda^{0}}_{i}=(v^{i}/c)/  \sqrt{1-v^{2}/c^{2}} ,
\nonumber \\
{\Lambda^{i}}_{j}&=&\delta_{ij}+ v^{i} v^{j}
\frac{(1/\sqrt{1-v^{2}/c^{2}}-1)}{v^{2}}. \nonumber \\
\end{eqnarray}

For the bradyon, the world-velocity ${\cal U}^{\mu}$ becomes the usual velocity
four-vector
${\it u}^{\mu}(\tau)$, which is tangential to the trajectory with increasing
$\tau$. In this case, ${\it u}^{\mu}$ is time-like, {\it i.e.}, $u^{\mu}
u_{\mu}
=c^{2} >0$. On the other hand, the corresponding quantity $t^{\mu}$ for the
tachyon, which is real-valued, being defined as follows
\begin{eqnarray}
t^{\mu} \equiv  i~{\cal U}^{\mu} \nonumber
   &=&\left( \frac{c~ dt}{d \lambda},
              \frac{d {\bf x}}{d \lambda} \right) \nonumber \\
&=&\left( \frac{c}{\sqrt{u^{2}/c^{2}-1}},\frac{{\bf u}}{\sqrt{u^{2}/c^{2}-1}}
 \right)
\end{eqnarray}
is the tangential to the world trajectory with increasing $\lambda$, and is
space-like, {\it i.e.}, $t^{\mu}t_{\mu }=-c^{2} <0$. The
transformation laws for these two distinct cases are given by
\begin{eqnarray}
{\it u'}^{\mu}&=&{\Lambda^{\mu}}_{\nu} {\it u}^{\nu}, \nonumber \\
t'^{\mu}&=&~ \frac{1 }{\mbox{sign}\{ 1 - {\bf u}\cdot {\bf v} / c^{2} \}}~
{\Lambda^{\mu}}_{\nu} t^{\nu}.
\end{eqnarray}

As an another independent kinematic world-quantity, we introduce
world-acceleration vector ${\cal A}^{\mu}$ as follows
\begin{eqnarray}
{\cal A}^{\mu} \equiv \frac{d {\cal U}^{\mu} }{d {\it w}}.
\end{eqnarray}
Then due to the pseudo-vector property of ${\cal U}^{\mu}$ and pseudo-scalar
property of ${\it w}$, the world acceleration ${\cal A}^{\mu}$ becomes the
four-vector, $\it{i.e}$, it transforms the same as $dx^{\mu}$ under
the Lorentz transformation, as usual. It's components are given by
\begin{eqnarray}
\cal{A}^{\mu} &=&\left( \frac{c~ d^{2}t}{d {\it w}^{2}},
              \frac{d^{2} {\bf x}}{d {\it w}^{2}} \right) \nonumber \\
&=&\left( \frac{{\bf u}\cdot {\bf a}}{c~ (1-u^{2}/c^{2})^{2}},~
\frac{{\bf a}}{1-u^{2}/c^{2}} +
\frac{({\bf u} \cdot {\bf a})  ~{\bf u}}{c^{2}~ (1-u^{2}/c^{2})^{2}}
 \right)
\end{eqnarray}
with the square magnitude of the four-vector
\begin{eqnarray}
{\cal A}^{\mu} {\cal A}_{\mu} &=&-\frac{a^{2} c^{2}
+({\bf a}\cdot {\bf u})^{2}-a^{2}u^{2}}{c^{2}(1-u^{2}/c^{2})^{3} }
\nonumber \\\\
\nonumber &=&-\frac{\{a^{2} c^{2} -a^{2} u^{2}~ \mbox{sin}^{2} \phi\} }
{c^{2}(1-u^{2}/c^{2})^{3} },
\end{eqnarray}
where $\phi$ is the angle between the three-velocity vector ${\bf u}$ and
three-acceleration vector ${\bf a}=d {\bf u}/d t$. And this world-acceleration
vector ${\cal A}^{\mu}$ is the vector which is normal to the tangential vector
${\cal U}_{\mu}$,
{\it i.e.},
\begin{eqnarray}
{\cal U}_{\mu}{\cal A}^{\mu}={\cal U}_{\mu}~
                 \frac{d {\cal U}^{\mu} }{d {\it w}}=0
\end{eqnarray}
from the relation ${\cal U}_{\mu}(d{\cal U}^{\mu}/dw)=
(1/2)d({\cal U}_{\mu}{\cal U}^{\mu})/dw$
and
\begin{eqnarray}
{\cal U}_{\mu}{\cal U}^{\mu}=c^{2}.
\end{eqnarray}

For the bradyon, the world-acceleration vector coincides with the usual
four-acceleration vector ${\it a}^{\mu}$ as
\begin{eqnarray}
{\cal A}^{\mu} &=& \frac{d {\cal U}^{\mu} }{d {\it w}}
         = \frac{d {\it u}^{\mu} }{d \tau }
          \equiv  {\it a}^{\mu}
\end{eqnarray}
because ${\cal U}^{\mu}={\it u}^{\mu}$ and ${\it w}= \tau$ in this case. The
square of the ${\it a}^{\mu}$ is always negative definite due to the two facts:
\begin{eqnarray}
\mbox{ i) }&&~a^{2}c^{2}-a^{2}u^{2} \mbox{sin}^{2} \phi ~\geq ~a^{2}c^{2}-
            a^{2}u^{2} ~>~ 0, \nonumber \\
\mbox{ ii)}&& ~ (1-u^{2}/c^{2})^{3} ~>~ 0
\end{eqnarray}
,{\it i.e.}, $a^{\mu}$ is the space-like vector (Fig. 5).

On the other hand, for the tachyon, an interesting new phenomenon occurs. In
this case with the reasonable
definition of the acceleration vector
\begin{eqnarray}
a_{t}^{\mu} \equiv \frac{d t^{\mu}}{d \lambda}
           =- \frac{d {\cal U}^{\mu}}{d {\it w}}
           =-{\cal A}^{\mu}
\end{eqnarray}
since $t^{\mu}=i ~{\cal U}^{\mu}$ and $\lambda=-i~ {\it w}$ in this case, the
norm of $a_{t}^{\mu}$ is not always negative value due to the two facts:
\begin{eqnarray}
\mbox{ i) }&&~a^{2}c^{2}-a^{2}u^{2} \mbox{sin}^{2} \phi ~\frac{<}{>} ~0
{}~~~\mbox{depending~on}~~~ \mbox{sin}^{2}\phi ~\frac{<}{>}
{}~\frac{c^{2}}{u^{2}},
\nonumber \\
\\ \nonumber
\mbox{ ii)}&& ~ (1-u^{2}/c^{2})^{3} ~<~ 0.
\end{eqnarray}
Rather $a^{\mu}_{t}$ is space-like or time-like or light-like depending on the
value of angle $\phi$ for a given velocity ${\bf u}$ (Fig. 6).\\

{\bf C}. {\bf Least action approach to the tachyon dynamics}\\

In the least action approach one fundamentally assume that one can define a
world-scalar action by which we obtain the covariant equations of
motion of the interested objects by applying an invariant principle, {\it
i.e.},
the least action principle. Therefore, it is essential to find the Lorentz
invariant combination called the action, which is appropriate for the system
considered, or introduce some new parameters and variables such that the action
is Lorentz invariant to proceed this approach if there is no non-trivial form
of
the Lorentz invariant action with the given dynamical variables.

Now, since for the free objects the only dynamical variables are
${\cal U}^{\mu}$, the possible candidate for the invariant combinations
are ${\cal U}^{\mu} ~{\cal U}_{\mu},~( {\cal U}^{\mu} ~{\cal U}_{\mu} )^{2}$,
..., etc. . However, unfortunately only these terms are not appropriate for the
description of the dynamics because these terms are nothing
but the power of non-dynamical universal constant $c$ from the relation (59).
Since these are all the possible Lorentz invariant combinations from the
dynamical variable ${\cal U}^{\mu}$, we must
consider the other possibilities to construct the Lorentz invariant action.
To this end we first consider the world-parameter
$w$. However, since this is not Lorentz invariant but pseudo-invariant,
we must introduce some pseudo-invariant coefficient to cancel out the
additional sign-factor which comes from the transformation law of $w$, and
hence make their
product Lorentz invariant.

Let us define the Lorentz invariant world-action ${\cal J}$ describing the
tachyon as well as the bradyon formally, as
\begin{eqnarray}
{\cal J}[~w_{1},~w_{2}~] =-\int^{w_{2}}_{w_{1}} {\cal U}^{\mu}~ {\cal U}_{\mu}
K
               ~dw
                      =-c^{2} \int^{w_{2}}_{w_{1}} K ~dw
\end{eqnarray}
where $K$ has the mass dimension, and has pseudo-invariant property under the
Lorentz transformations (51) and (52), ${\it i.e.}$,
\begin{eqnarray}
K'&=&\mbox{sign}\left\{ 1 -\frac{ {\bf u}\cdot {\bf v} }{ c^{2}}
\right\} K \\ \nonumber
\end{eqnarray}
such that the product of $K$ and $dw$ is Lorentz invariant. We call $K$ as
`` the world-mass parameter ''. Moreover, note that the combination
$\int {\cal U}^{\mu}dx_{\mu}$ reduces also to the form of (64). This is our
fundamental assumption of the least action approach to the general
dynamics of the
free objects with the tachyon as well as the bradyon. When expressed by
the three-velocity ${\bf u}$ with the relation (40), the world-action becomes
\begin{eqnarray}
{\cal J}=-c^{2} \int^{t_{2}}_{t_{1}}K \sqrt{1-u^{2}/c^{2}} ~dt,
\end{eqnarray}
which explicitly shows the reparametrization invariance
$t \rightarrow \tilde{t} (t)$ for the monotonically increasing function
$\tilde{t} (t)$ such that
\begin{eqnarray}
\left|\frac{d \tilde{t}}{dt} \right|= \frac{d \tilde{t}}{dt}
\end{eqnarray}
is satisfied for all the time $t$. With the help of the reparametrization
invariance the least action principle can be also applied to the tachyon
without any  ambiguity as well as the bradyon. For the bradyon, the usual
(virtual) variational method to derive the equations of motion with the fixed
time
$t$ is well defined for any bradyonic trajectory since the time interval
$\Delta t=t_{A}-t_{B}$ for any two events $A$ and $B$ along the world-line
of the object is always non-zero such that the derivatives by ``$dt$'' have no
ambiguities.

By defining the Lagrangian $L(t)$ for the bradyon with the physical
time $t$ as
\begin{eqnarray}
{\cal J}[~t_{1},~t_{2}~] =\int^{t_{2}}_{t_{1}} L(t)~ dt
\end{eqnarray}
with
\begin{eqnarray}
L(t)=-m_{0}c^{2}  \sqrt{1-u^{2}/c^{2}}
\end{eqnarray}
and the canonical linear momentum ${\bf p}(t)$ as
\begin{eqnarray}
{\bf p}(t) \equiv \frac{\partial ~L(t) }{\partial ~(d{\bf x}/dt) }
           = \frac{\partial ~L(t)}{\partial ~{\bf u}}
           =\frac{m_{0} {\bf u}}{\sqrt{1-u^{2}/c^{2}}},
\end{eqnarray}
the equation of motion becomes
\begin{eqnarray}
\frac{d {\bf p}}{dt}=0
\end{eqnarray}
from the least action condition
\begin{eqnarray}
0=\delta {\cal J}(~t_{1},~t_{2}~)&=& \int^{t_{2}}_{t_{1}}\delta L(t) ~dt
\nonumber \\\\
&=&\int^{t_{2}}_{t_{1}} \left[ \frac{\partial ~L}
{\partial ~{\bf x}}-\frac{d}{dt} \left( \frac{\partial ~L}{\partial
         ~(d {\bf x}/dt )} \right)
\right] \cdot \delta {\bf x}~ dt  \nonumber
\end{eqnarray}
with the auxiliary condition
\begin{eqnarray}
\delta {\bf x}(t_{1}) =  \delta {\bf x}(t_{2}) =0
\end{eqnarray}
under the fixed time (virtual) variation (Fig. 7). Here we designate the
world-mass parameter by the Lorentz invariant mass $m_{0}$ since in this case
$ K $ is not the pseudo-scalar any more but only a Lorentz scalar.
The corresponding Hamiltonian becomes
\begin{eqnarray}
H(t) \equiv {\bf p} \cdot {\bf u} -L
  =\frac{ m_{0} c^{2} }{ \sqrt{ 1-u^{2}/c^{2} } },
\end{eqnarray}
which is also constant in time for the system when
\begin{eqnarray}
\frac{\partial ~L}{\partial ~t} =0
\end{eqnarray}
is satisfied, which is actually the case of ours. Furthermore, since the action
(68) and the least action principle (72) are Lorentz invariant, the
conservation
law of the momentum ${\bf p}$ and energy $E$ are also satisfied in all
uniformly
moving systems as in the case of the previous section. For many objects case,
the conserved ones are found to be the total momentum  and energy.

On the other hand, for the tachyon this approach is more involved, {\it i.e.},
the usual variational
method to derive the equations of motion with fixed time $t$ has some problem
in this case. This is essentially due to the time interval $\Delta
t=t_{B}-t_{A}$ between two events $A$(leaving of the tachyon) and $B$(arriving
of
the tachyon) along the world-line of the object always vanish for the infinite
velocity tachyon, which is called {\it transient} tachyon [8], such that there
are some ambiguities in deriving the equation
of motion. Furthermore, since any tachyon can be made to have the infinite
velocity by choosing an appropriate system of coordinates, it seems that this
problem can not be avoided for any tachyon (Fig. 8). However, this problem can
be overcome by using the reparametrization invariance.

To see this, we note that the world-action (66) parameterized by the tachyon's
world-parameter $\lambda$ of Eqs. (43) and (44) becomes
\begin{eqnarray}
{\cal J}&=&c^{2} \int^{t_{2}}_{t_{1}} \kappa \sqrt{ \frac{1}{c^{2}}
        \left( \frac{d {\bf x}}{dt} \right)^{2}-1 }~~dt \nonumber \\
        \nonumber \\
&=&c^{2} \int^{t_{2}}_{t_{1}} \kappa \sqrt{ \frac{1}{c^{2}}
    \left(\frac{d {\bf x}}{d \lambda}\right)^{2}-\left(\frac{dt}{d \lambda }
          \right)^{2} }
   ~ \left| \frac{d \lambda}{dt}\right| ~dt  \\ \nonumber \\
&=&c^{2} \int^{\lambda_{2}}_{\lambda_{1}} \kappa \sqrt{ \frac{1}{c^{2}}
   \left(\frac{d {\bf x}}{d \lambda}\right)^{2}-\left(\frac{dt}{d \lambda }
         \right)^{2} }~~d \lambda ,
    \nonumber
\end{eqnarray}
where we have introduced the mass parameter $\kappa \equiv -i K$, which should
be real-valued in order that the action is real-valued [22], and
has the same transformation law as that of $K$, {\it i.e.,}
\begin{eqnarray}
\kappa'&=&\mbox{sign}\left\{ 1 -\frac{ {\bf u}\cdot {\bf v} }{ c^{2}}
\right\} \kappa. \\ \nonumber
\end{eqnarray}
Here, we have used the fact [23]
\begin{eqnarray}
\frac{d \lambda}{d t}=\sqrt{u^{2}/c^{2}-1}~~ > 0.
\end{eqnarray}

Now, in this parameterization there is no singular case of $\Delta \lambda
\equiv 0$ because $\Delta \lambda $ has an invariant magnitude and $\Delta
\lambda $ is non-zero for any tachyon, including the transient tachyon as well,
for any two different points along the
world-line. The singular point of $\Delta t$ at $u=\infty$ is transferred to
the singular point of $\Delta \lambda$ at $u=c$, {\it i.e.}, the case
of luxon such that there is no problem for the tachyon only case. Now, with
the help of this property, the least action principle can be also applied in
this case by adopting the fixed $\lambda$ (virtual)
variations (Fig. 9) without any ambiguity .

By defining the Lagrangian $L(\lambda)$ with the parameter $\lambda$ as
\begin{eqnarray}
{\cal J}[~\lambda_{1},~\lambda_{2}~]
 =\int^{\lambda_{2}}_{\lambda_{1}} L(\lambda) ~d \lambda
\end{eqnarray}
with
\begin{eqnarray}
L( \lambda)= \kappa c^{2}
\sqrt{ \frac{1}{c^{2}}
   \left(\frac{d {\bf x}}{d \lambda}\right)^{2}-\left(\frac{dt}{d \lambda }
   \right)^{2} }~~d \lambda ,
\end{eqnarray}
the equations of motion become
\begin{eqnarray}
0&=&  \frac{\partial ~L}
{\partial ~t}-\frac{d}{d \lambda} \left[ \frac{\partial ~L}
{\partial ~(d t/d \lambda)} \right] =-\frac{d}{d \lambda}
\left[ \frac{\partial ~L}{\partial ~(d t/d \lambda)} \right] ,\\ \nonumber \\
0&=&  \frac{\partial ~L}
{\partial ~{\bf x}}-\frac{d}{d \lambda} \left[ \frac{\partial ~L}
{\partial ~(d {\bf x}/d \lambda)} \right] =-\frac{d}{d \lambda}
\left[ \frac{\partial~ L}{\partial ~(d {\bf x}/d \lambda)} \right]
\end{eqnarray}
from the least action condition
\begin{eqnarray}
0&=&\delta_{[ \lambda ]}{\cal J}[~\lambda_{1},~\lambda_{2}~]
 =\int^{\lambda_{2}}_{\lambda_{1}} \delta_{[ \lambda ]} L(\lambda)~ d \lambda
 \nonumber \\ \nonumber \\
 &=&\int^{\lambda_{2}}_{\lambda_{1}}
 \left\{
 \left[
  \frac{\partial ~L}
{\partial ~t}-\frac{d}{d \lambda} \left[ \frac{\partial ~L}
{\partial~ (d t/d \lambda)} \right]
                  \right]  \delta_{[ \lambda ]} t
 +\left[
  \frac{\partial ~L}
{\partial ~{\bf x}}-\frac{d}{d \lambda} \left[ \frac{\partial ~L}
{\partial ~(d {\bf x}/d \lambda)}\right]
                   \right]  \cdot \delta_{[ \lambda ]} {\bf x}
\right\}      d \lambda
\end{eqnarray}
with the auxiliary conditions
\begin{eqnarray}
\delta_{[ \lambda ]}~ x^{\mu}(\lambda_{1}) =\delta_{[ \lambda ]}
{}~x^{\mu}(\lambda_{2})=0,
\end{eqnarray}
and where $\delta_{[ \lambda ]}$ represents the variation with fixed $\lambda$.
In this case the quantity
\begin{eqnarray}
{\bf \Pi}(\lambda) \equiv \frac{\partial ~L}{\partial ~(d {\bf x}/d \lambda) }
=\frac{\kappa ~(d {\bf x}/d \lambda) }
{\sqrt{( d {\bf x}/c~d \lambda)^{2}  - (dt/d \lambda )^{2} }  }
\end{eqnarray}
can be considered as the canonical momentum, which is conserved with respects
to the evolution of parameter $\lambda$ in all the uniformly moving systems
because the
conservation law (81) is satisfied in all uniformly moving systems due to the
invariance of the world-action (79) and the least action principle (83). When
expressed in the three-vector ${\bf u}$, ${\bf \Pi}(\lambda)$ becomes
\begin{eqnarray}
{\bf \Pi}(\lambda)
=\frac{\kappa ~ {\bf u} }
{\sqrt{ u^{2}/c^{2}  - 1 } }
 \equiv {\bf p}(t),
\end{eqnarray}
which is formally the same result as in the previous section.

Furthermore, the quantity
\begin{eqnarray}
- \Pi^{0}(\lambda) \equiv - \frac{\partial ~L}{\partial ~(d t/d \lambda) }
=\frac{\kappa c^{2}~ (d t/d \lambda) }
{\sqrt{( d {\bf x}/c ~d \lambda)^{2}  - (dt/d \lambda )^{2} } }
\end{eqnarray}
can be considered, with respect to the evolution of the parameter $\lambda$, as
the energy, which is conserved invariantly like as the energy of the bradyon.
When expressed in the three-vector ${\bf u}$, $-\Pi^{0}(\lambda)$ becomes
\begin{eqnarray}
- \Pi^{0}(\lambda)
=\frac{\kappa c^{2}  }
{\sqrt{u^{2}/c^{2} - 1 } }
 \equiv E(t),
\end{eqnarray}
which is also formally the same result as in the previous section. However,
actually the replacement of the functions of $\lambda$, ${\bf \Pi}(\lambda)$
and
$-\Pi^{0} (\lambda)$ with the functions of $t$, ${\bf p}(t)$, and $E(t)$ is
non-trivial matter because $\lambda$ may not be a function of $t$ in the
mathematical rigor although $t$ is the function of $\lambda$ generally by
having in mind the infinite velocity tachyon (Fig. 10). Hence the replacement
of Eqs. (86) and (88) should be understood as
${\bf p}(t) \equiv{\bf p}(t(\lambda))$ and $ E(t) \equiv E(t(\lambda))$.
Moreover, in order to prove that they are also the conserved quantities,
{\it i.e.,} constant, with respect to the physical time
parameter $t$ `` without any ambiguity '', let us consider the change of
${\bf p}(~t(\lambda)~)$ and $E(~t( \lambda)~)$ in a small interval of $\lambda$
along the world-line
\begin{eqnarray}
d{\bf p}(~t(\lambda)~)&=&{\bf p}(~t_{2}(\lambda_{2})~)-{\bf p}(~t_{1}(
      \lambda_{1})~), \nonumber \\
dE(~t(\lambda)~)&=&E(~t_{2}(\lambda_{2})~)-E(~t_{1}(\lambda_{1})~),
\end{eqnarray}
where $\lambda_{2}=\lambda_{1}+d \lambda$. Then, by using the equations of
motion
(82) and (83) we can easily see that
\begin{eqnarray}
d{\bf p}(~t(\lambda)~)&=&\left(\frac{d {\bf p}}{d \lambda} \right) d \lambda
=0,
\nonumber \\\\
dE(~t(\lambda)~)&=&\left(\frac{d E}{d \lambda} \right) d \lambda =0
\nonumber
\end{eqnarray}
for any $d \lambda$ and hence for any $dt$ implying that ${\bf p}$ and
$E$ are really the conserved ones with respect to the time $t$ along the
world-line for the transient tachyon also as well as the other tachyons.

Therefore, in this section we have shown that the least action principle can
be also applied to the tachyon, and moreover
by introducing this principle we have reproduced the essential
results for the tachyon of the
previous section, {\it i.e.,} the energy and momentum formula and the mass
transformation formula only by considering the single object case and
without consideration of the many objects collision processes, which has been
essential to derive the results uniquely. This corresponds to a merit of the
least action approach. However, for the determination of the zero-point energy
of the point object in the previous section, it seems that the elementary
approach of that section is more clear and simple, and so this will not be
reconsidered in this section.
Finally, we
note that the results of the tachyon as well as the bradyon have been obtained
independently on the existence of the luxon.\\

\begin{center}
{\large \bf IV. THEORY OF THE LUXON} \\
\end{center}

For the description of the point-like object moving with the light velocity,
{\it i.e.}, luxon, the analysis of the previous two sections can not be
directly applied to it, as a limit case of $lim~u \rightarrow c$ without
any ambiguity. This can be easily seen by considering, for example, the formula
(70) and (74) of the momentum and energy of the bradyon  for the case of
$lim~u \rightarrow c$ in all uniformly moving systems. In
this limit, both ${\bf p}$ and $E$ are ill-defined:\\

a) for non-zero
proper-mass luxon, only the infinite value of ${\bf p}$ and $E$ are allowed in
any systems. This luxon is not believed to be detected in our essentially
finite-mass bradyonic detectors, and hence is unphysical objects, and \\

b) for zero proper-mass luxon, both ${\bf p}$ and $E$ has the form of $0/0$ and
hence they are also ill-defined unless we know the explicit
pattern of $m_{0} \rightarrow 0$ as $u \rightarrow c$ [24].\\

Furthermore,
since both ${\bf p}$ and $E$ are ill-defined, their any  interrelations are
also
ill-defined such that the usual formula of the luxon
\begin{eqnarray}
E=pc
\end{eqnarray}
can not be justified in `` a purely mechanical sense '' although it is
satisfied
for all examples of luxon as far as we know [25]. These facts are also true for
the limit $u \rightarrow c$ from the tachyon formulas. Hence, both the bradyon
and tachyon have ill-defined limits of $u \rightarrow c$. Actually this fact is
not surprising because all derivations of the previous sections II and III
can not be justified for the case of the luxon with mathematical rigor. Hence,
for the consistent description of the luxon, an independent theory is needed.\\

{\bf Least action approach to the luxon dynamics}\\

With the help of the least action principle, we can naturally and easily
construct the theory of luxon without any ambiguity. To this end, we first note
that the world-parameter $w$ is not appropriate for the description of the
world-trajectory of the luxon because \\

i) $\Delta w=\Delta t~ \sqrt{ 1-u^{2}/c^{2} }=\Delta t \cdot 0=0$ for
any time interval $\Delta t$ with $|\Delta t| < \infty$ and \\

ii) $\Delta w=\Delta t \cdot 0$ has ambiguous value for $|\Delta t|=\infty$.\\

Here $\Delta w \equiv w_{B}-w_{A},~ \Delta t \equiv
t_{B}-t_{A}$ are differences of their values for any two events
$A$(leaving of luxon) and $B$(arriving of luxon) along the world-line.
Furthermore, due to this fact we note secondly that from the previous formula
(49) the world-velocity
${\cal U }^{\mu}$ of the luxon has the indefinite values
\begin{eqnarray}
{\cal U}^{\mu} \equiv \frac{d x^{\mu}}{d w}
=\left( \frac{c}{0}, \frac{ c {\hat k}}{0} \right),
\end{eqnarray}
where ${\hat k}$ is the unit propagation vector
and hence ${\cal U}^{\mu}$ is not appropriate for the description of the
dynamics of the luxon having the well-defined definite velocity by
definition.

In order to avoid this problem we introduce a real-valued parameter $\eta$
describing the trajectory of the luxon $x^{\mu}=x^{\mu}(\eta)$ by
\begin{eqnarray}
\frac{d x^{\mu}}{d \eta}  \frac{d x_{\mu}}{d \eta} &=&\frac{d x^{\mu}}{d t}
\frac{d x_{\mu}}{d t}  ~ \left( \frac{d t}{d \eta} \right)^{2} \nonumber \\
    &=&(c^{2}-u^{2}) ~\left( \frac{d t}{d \eta} \right)^{2} \\
    &=&0  \nonumber
\end{eqnarray}
with the condition
\begin{eqnarray}
\left( \frac{d t}{d \eta} \right)^{2} ~ < ~\infty
\end{eqnarray}
such that Eq. (93) characterizes the case of the luxon. Then, in order that
these
defining relations are Lorentz-covariantly valid, some
restrictions on $\eta$ are needed. By defining the Lorentz transformation of
$\eta$ as $\eta'=\eta' (\eta)$, {\it i.e.}, the transformation law of $\eta$
under the Lorentz transformation can be expressed as is independent of the
space
and time coordinates of the objects, the relation (93) is covariantly satisfied
as follows
\begin{eqnarray}
\frac{d {x^{\mu}}'}{d \eta '}  \frac{d {x_{\mu}}'}{d \eta '}
=\frac{d x^{\mu}}{d \eta}    \frac{d x_{\mu}}{d \eta}
{}~\left( \frac{d \eta}{d \eta '} \right)^{2} =0,
\end{eqnarray}
only if
\begin{eqnarray}
\left( \frac{d \eta}{d \eta '} \right)^{2} ~<~ \infty
\end{eqnarray}
,{\it i.e.}, the Lorentz transformation $\eta'=\eta'(\eta)$ is a non-singular
transformation. Furthermore, the condition (94) is also covariantly satisfied
as
follows
\begin{eqnarray}
 \left( \frac{d t'}{d \eta '} \right)^{2}&=&\left[
      \gamma ~\left( \frac{dt}{d \eta '}
  -\frac{{\bf v}}{c^{2}} \cdot \frac{d {\bf x}}{d \eta'} \right) ~\right] ^{2}
  \nonumber \\
   &=&{\gamma}^{2} \left(1-\frac{{\bf v}}{c^{2}} \cdot {\bf c}
   \right)^{2}~ \left( \frac{d t}{d \eta } \right)^{2}   ~
   \left( \frac{d \eta}{d \eta ' } \right)^{2} ~<~ \infty ,
\end{eqnarray}
only if Eq. (96) is satisfied. Hence, only if we consider non-singular
transformation $\eta '=\eta' (\eta)$ under the Lorentz transformation, the
defining relations (93) and (94) can be satisfied covariantly. Now, by
defining the null-velocity vector of the luxon $l^{\mu}$ as
\begin{eqnarray}
l^{\mu} \equiv \frac{d x^{\mu}}{d \eta},
\end{eqnarray}
the possible candidate for the invariant combinations from this velocity
vector are $l^{\mu}l_{\mu}$,
$(l^{\mu}l_{\mu})^{2}$, ..., etc.. Although it seems that the property of
\begin{eqnarray}
l^{\mu}l_{\mu} =0
\end{eqnarray}
make these combinations inappropriate as the building blocks of the
Lorentz-invariant action at first, this is not the case because this equation
is
a dynamical equation in contrast to the non-dynamical mathematical identity
(59) of the bradyon and the tachyon: Eq.(99) does express that the velocity of
the luxon is always the light velocity which being the only mechanical property
of the luxon [26]. Hence the action of the luxon, if it exists, should provide
the
equation of motion (93) or (99).

To realize this scenario we introduce a variable $\xi (\eta)$ into the action
of the luxon by
\begin{eqnarray}
{\cal J}[~\eta_{1},~ \eta_{2}~]=-\frac{1}{2} \int _{\eta_{1}}^{\eta_{2}}
l^{\mu} l_{\mu}~ \xi ~d \eta
\end{eqnarray}
analogous to $K$ in Eq.(64), but now $\xi$ is not the pseudo-scalar or scalar
in
general rather it's transformation under the Lorentz transformation is the same
as ``$d \eta$'',
{\it i.e.},
\begin{eqnarray}
\xi(\eta) \rightarrow \xi '(\eta')=\left( \frac{d \eta'}{d \eta} \right)
\xi (\eta) ~~
\mbox{ when}~~
d \eta  \rightarrow d \eta' =\left( \frac{d \eta '}{d \eta} \right) d \eta
\end{eqnarray}
in order that the action (100) is Lorentz invariant. Here the factor 1/2 in
front of the action is introduced for convenience. Note that although by
varying ${\cal J}$ with respect to $\xi$ we can recover the equation of motion
of luxon (93) or (99), the variable $\xi$ should not be considered as an
auxiliary field as in the case of usual `${\it einbein}$' formulation [27].
This means that the equation of motion with respect to variation of $\xi$
should not be implemented back into the action (100) as in the case of
auxiliary
field: if it is implemented to (100), the action becomes vanishing, which does
not show any dynamics of the luxon. So we emphasize that the new variable
$\xi$ is not an auxiliary one although there is no derivatives of $\xi$ with
respect to $\eta$ in our formulation [28]. Moreover, we note that the Lorentz
transformation of $\xi$, Eq.(101) implies the invariance of the action (100)
under the reparametrization
\begin{eqnarray}
\eta \rightarrow \tilde{\eta} (\eta)
\end{eqnarray}
if the variable $\xi$ does not distinguish the variation of $\eta$ under the
Lorentz transformation and that of under the reparametrization, {\it i.e.},
\begin{eqnarray}
\xi (\eta) \rightarrow \tilde{\xi}(\tilde{\eta})
=\left( \frac{d \tilde{\eta}}{d \eta} \right) \xi (\eta).
\end{eqnarray}
Note that this assumption is plausible since $\xi$ is the function of only
$\eta$.
Hence, the Lorentz invariance implies the reparametrization invariance in this
case.

Now let's consider their all equations of motion. By defining the Lagrangian
$L(\eta)$ of the luxon with the parameter $\eta$ as
\begin{eqnarray}
{\cal J}[~\eta_{1},~ \eta_{2}~]=\int _{\eta_{1}}^{\eta_{2}} L(\eta) ~d \eta
\end{eqnarray}
with
\begin{eqnarray}
L(\eta)=-\frac{1}{2} \xi(\eta) ~l^{\mu} l_{\mu}
=-\frac{1}{2} \xi(\eta) \left[ \left(\frac{c dt}{d \eta} \right)^{2}
- \left(\frac{ d{\bf x}}{d \eta} \right)^{2}   \right],
\end{eqnarray}
the equations of motion become
\begin{eqnarray}
0&=&  \frac{\partial ~L}
{\partial ~t}-\frac{d}{d \eta} \left[ \frac{\partial ~L}
{\partial ~(d t/d \eta)} \right] =-\frac{d}{d \eta}
\left[ \frac{\partial ~L}{\partial ~(d t/d \eta)} \right], \\  \nonumber \\
0&=&  \frac{\partial ~L}
{\partial ~{\bf x}}-\frac{d}{d \eta} \left[ \frac{\partial ~L}
{\partial ~(d {\bf x}/d \eta)} \right] =-\frac{d}{d \eta}
\left[ \frac{\partial~ L}{\partial ~(d {\bf x}/d \eta)} \right]
\end{eqnarray}
from the least action condition
\begin{eqnarray}
0&=&\delta_{[ \eta ]}{\cal J}[~\eta_{1},~\eta_{2}~]
 =\int^{\eta_{2}}_{\eta_{1}} \delta_{[ \eta ]} L(\eta) ~d \eta
 \nonumber \\ \nonumber \\
 &=&\int^{\eta_{2}}_{\eta_{1}}
 \left\{
 \left[
  \frac{\partial ~L}
{\partial ~t}-\frac{d}{d \eta} \left[ \frac{\partial ~L}
{\partial ~(d t/d \eta)} \right]
                  \right]  \delta_{[ \eta ]} t
 +\left[
  \frac{\partial L}
{\partial {\bf x}}-\frac{d}{d \eta} \left[ \frac{\partial~ L}
{\partial ~(d {\bf x}/d \eta)}\right]
                   \right]  \cdot \delta_{[ \eta ]} {\bf x}
\right\}      d \eta
\end{eqnarray}
with the auxiliary conditions
\begin{eqnarray}
\delta_{[ \eta ]}~ x^{\mu}(\eta_{1}) =  \delta_{[ \eta ]}~ x^{\mu}(\eta_{2})
=0.
\end{eqnarray}
In this case the quantity
\begin{eqnarray}
{\bf \Pi}(\eta) \equiv \frac{\partial ~L}{\partial ~(d {\bf x}/d \eta) }
=\xi ~\frac{d {\bf x}}{d \eta}
\end{eqnarray}
can be considered as the canonical momentum, which is conserved with respect
to the evolution of parameter $\eta$ in uniformly moving systems from the same
reason as in the case of the tachyon. When expressed in the three-vector
${\bf u}=c \hat{k}$ with unit propagation vector $\hat{k}$, ${\bf \Pi} (\eta)$
becomes
\begin{eqnarray}
{\bf \Pi}(\eta) =\left( \frac{dt}{d \eta}\right) \xi c \hat{k},
\end{eqnarray}
which is invariant under the reparametrization (102) and (103).

On the other hand, the quantity
\begin{eqnarray}
-\Pi_{0}(\eta) \equiv -\frac{\partial ~L}{\partial~(dt/d \eta)}
=\left(\frac{dt}{d \eta}\right) \xi c^{2}
\end{eqnarray}
can also be considered as the conserved energy with respects to the
evolution of parameter $\eta$ invariantly, and also is invariant under the
reparametrization (102) and (103). Furthermore, since $\eta$ can always be
considered as a function of $t$ due to the condition of (94), it is trivial to
represent the functions of $\eta$, ${\bf \Pi}(\eta)$ and $\Pi_{0}(\eta)$ as the
functions of $t$ as
\begin{eqnarray}
{\bf \Pi}(\eta) \equiv {\bf p}(t),~~-\Pi_{0}(\eta) \equiv E(t)
\end{eqnarray}
such that
\begin{eqnarray}
\frac{d {\bf p}(t)}{d t} =0,~~\frac{d E(t)}{d t} =0
\end{eqnarray}
are satisfied, {\it i.e.,} ${\bf p}(t)$ and $E(t)$ are conserved quantities
with respect to the evolution of the time $t$. Note that we have now expressed
the dynamical quantities of the
luxon without any ambiguity by the function
\begin{eqnarray}
(dt/d \eta)~\xi,
\end{eqnarray}
which is the reparametrization invariant, and by the propagation vector
$\hat{k}$. Hence the essential role of $\xi$ is that it acts as a new dynamical
parameter or variable when combined with the quantity $(dt/d \eta)$
for the unambiguous description of the system of the luxon.

Furthermore, it is easy to see that the momentum and energy formulas
(110)-(112)
really show the usual energy-momentum relation for the luxon
\begin{eqnarray}
E^{2}-p^{2}c^{2} =0,
\end{eqnarray}
and the transformation under the Lorentz transformation
\begin{eqnarray}
{\bf p}'&=&\xi ' \frac{d {\bf x}'}{d \eta '}=\xi  \frac{d {\bf x}'}{d \eta }
\nonumber \\ \nonumber \\ \nonumber
        &=&\left[ 1 + ( \gamma- 1 ) \frac{  {\bf v} } {v^{2} }  {\bf v}~ \cdot
{}~
        \right]
         \xi  \frac{d {\bf x}}{d \eta}
      -\gamma \frac{{\bf v}}{c^{2}}~
      c^{2} \xi  \frac{d t}{d \eta } \\ \nonumber  \\ \nonumber
       &=&\left[ 1 + ( \gamma- 1 ) \frac{  {\bf v} } {v^{2} }  {\bf v}~ \cdot ~
       \right]
       {\bf p}
      -\gamma \frac{{\bf v}}{c^{2}} E,   \nonumber
      \\ \\ \nonumber
E'&=& c^{2} \xi ' \frac{d t'}{d \eta '}=c^{2} \xi \frac{d t'}{d \eta }
      \\ \nonumber \\ \nonumber
  &=&\gamma~\left( c^{2} \xi \frac{d t}{d \eta }-
  \xi  \frac{d {\bf x}}{d \eta } \cdot {\bf v} \right)  \\ \nonumber \\
  \nonumber
  &=&\gamma~\left( E-{\bf p} \cdot {\bf v} \right)
       \\ \nonumber
\end{eqnarray}
according to the Lorentz transformation (101) of $\eta$ and $\xi$ without
any ambiguity, which has been impossible with mathematical rigor due to the
ill-defined nature of the energy and momentum of the bradyon and tachyon for
the luxon limit.

Now, with the help of this energy and momentum transformation, we can easily
prove by considering the collision process of the arbitrary number, and kind
of the luxons should satisfy the same constraints as in the case of
the tachyon and bradyon of Sec. II
if we maintain the definition of the energy and
momentum as the conserved quantities for all the systems of coordinates and
all the possible process of the collision. Especially, for the case of the
photon it's zero-point energy should be zero due to the fact that it is
clearly `` the strictly neutral particles ''.  Furthermore, due to the
transformation (117) we
can also define the four-momentum
\begin{eqnarray}
p^{\mu}=(E,~ c{\bf p})=\left( c^{2} \xi \frac{d t}{d \eta} ,
{}~ c \xi \frac{d {\bf x}}{d \eta} \right) =c \xi \frac{d x^{\mu}}{d \eta},
\end{eqnarray}
which is the light-like, ${\it i.e.,}$
\begin{eqnarray}
p^{\mu} p_{\mu}={p^{\mu}}'{p_{\mu}}'=0
\end{eqnarray}
due to Eq. (116), and hence conservation law (114) can be expressed as a
covariant form
\begin{eqnarray}
\frac{d p^{\mu}}{d t}=0.
\end{eqnarray}

It is interesting to note that the luxon exists only as a quantum object in our
universe as far as we know. In this case, with the help of the Einstein
relation
for the photon or neutrino
\begin{eqnarray}
E=\hbar c k, ~~{\bf p}= \hbar k \hat{k},
\end{eqnarray}
the function (115) is found to be
\begin{eqnarray}
(dt/d \eta) ~\xi =\hbar k.
\end{eqnarray}
\\

\begin{center}
{\large \bf V. SUMMARY AND CONCLUDING REMARKS}
\end{center}

In this paper we have presented the classical foundation of the relativistic
dynamics including the tachyon. An anomalous sign-factor in the transformation
of
${\sqrt{1-u^{2}/c^{2}}}$ under the Lorentz transformation, which is unimportant
for the bradyon and has been always missed in the usual formulation of the
tachyon, has now a crucial role in the dynamics of the tachyon. Due to this
novel sign-factor, it is found from the consideration of the collision process
that the proper-mass of the tachyon is not an absolute quantity in contrast to
that
of the bradyon, but transforms as $\kappa'= \mbox{sign}\{1-{\bf u \cdot
v}/c^{2}
\}
\kappa$. Furthermore, we
have shown that the tachyon's mass has a definite sign for a given speed in a
uniformly moving system, but has the sign changed depending on the tachyon's
velocity and the relative motion of the systems $S$ and $S'$.

On the other hand, by defining the momentum and energy as the conserved
quantities of the form Eq. (22) and (24) in all uniformly moving systems, we
have shown that the zero-point energy $\epsilon_{a}$ for any kind of the
objects of the both tachyon and bradyon, which has been known as the
undetermined constant, should satisfy some constraints ,{\it i.e.},
$\epsilon_{a}=0$ for the strictly neutral particles and
$\epsilon_{a~\mbox{\scriptsize particle}}=-\epsilon_{a~\mbox{\scriptsize
antiparticle}}$ in other cases
however remainly undetermined for the latter case for consistency.
Furthermore, especially for the case of $\epsilon_{a} \equiv 0$ for all $a$ as
in the usual conventions, it is found that the energy and momentum
for the tachyon as well as the bradyon satisfy the usual four-vector
transformation such that the energy for the tachyon does not have the invariant
sign of the energy, but the sign can be changed depending on the object's
velocity and the relative motion of the two systems of coordinates $S$ and $S'$
in contrast to that of the bradyon. However, we have noted that this
transformation
formulas can not be obtained from the usual formulation of the tachyon in
contrast to the usual belief such that the original
motivation of the BDSF's reinterpretation principle, if it applies to our
Nature, can not be found in the
usual formulation but only in our new formulation of the tachyon.

As an alternative approach we have also presented the least action approach to
the
tachyon dynamics. Although it is not clear whether this approach can be also
applied
to the tachyon since the situation of $\Delta t \equiv 0$ for the time
interval along the world-line can not be avoided for any tachyon such that
there are some ambiguities in deriving the equations of motion. However, we
have
shown that this can be really also applied to the tachyon with the help of the
reparametrization invariance of the action, and the essential results for the
tachyon of the elementary approach of Sec. II can be also rederived in this
approach.

Furthermore, we have shown that an unambiguous description of the luxon
dynamics
is possible with the help of the least action approach, which has not been
clear
in the approach of Sec. II. In this approach, to this end a new dynamical
variable $\xi$ with the parameter $\eta$ of the luxon's world-line should have
been introduced to the action and with the help of these new variables, the
usual energy-momentum relation for the luxon (116) is proved without any
ambiguity.
Furthermore, it is shown that in this approach the energy and momentum of the
luxon also satisfy the usual transformation like as the tachyon and bradyon,
which has not been proved in the usual classical mechanical formulation.
Moreover, with the help of this energy and momentum transformation under the
Lorentz transformation, the zero-point energy for any kind of the luxons should
also satisfy the same constraints as in the case of the tachyon and bradyon of
Sec. II, if we maintain the definition of the energy and momentum
as the conserved quantities for all the systems of coordinates and all the
possible process of the collision like as the case of the tachyon and bradyon.
Especially, for the case of the photon it's zero-point energy should be zero
due to the fact that it is clearly `` the strictly neutral particles ''.
Moreover, we have noted that the luxon exists only as a quantum object as far
as
we know, and hence in that case the newly introduced variables $\xi$ combined
with $\eta$ are related with the variables of the ${\it wavicle}$.

Irrespective of these results, there are several problems for the tachyon,
which
should be clarified, as a) the problem of the causality violation [2,3,4],
b) the problem of the \v{C}erenkov effect in the vacuum [29,30], and
c) the problem of the quantization [2,9].

For the causality violation by noting that this effect is essentially due
to the interaction between the tachyon and bradyon [2,3,4], we are questioning
whether some conditions on the interaction modes, which obstruct the
measurement
of the effect, can be obtained from the detailed analysis of their interaction.
Or the usual concept of the space and time, which are defined by the measuring
rods and clocks at rest relatively to that system `` without references '' to
the motions, and the properties of the measured object may not be applied to
the
case of the tachyon, and be needed modification in order to recover the
causality in
this case also. On the other hand, in our opinion it is questionable whether
the principle of
causality can be considered as the fundamental principle of our Nature like as
the principles of the theory of relativity, and hence we might think that the
communication via the bradyon or luxon is a special case of the preservation of
causality but not true in general with the tachyon.

For the \v{C}erenkov effect in the vacuum, there has been no complete
agreement for the effect in relation mainly with the non-covariance of this
effect, {\it i.e.}, `` the transient tachyon problem '' [8,30], which states
that the transient charged tachyon, {\it i.e.}, zero energy (considered as the
lowest energy) tachyon does not radiate, but some radiation should be observed
in
other systems due to the non-invariance of the property of the transient
tachyon
such that the principle of relativity is violated if there exist \v{C}erenkov
effect in the vacuum. But, here we note that this may be not the case
essentially
because in the case of the tachyon the zero energy is not the lowest energy
state of it. Rather there are the lower states, {\it i.e.}, the negative energy
states such
that the zero energy tachyon may also radiate if there is the \v{C}erenkov
radiation for the positive energy tachyon. The reason for the inclusion of the
negative energy states into the energy range of the tachyon is that firstly,
the tachyons having different sign of the energy and hence of the mass are
connected by the Lorentz transformation, and secondly there is no energy gap in
the energy spectrum in contrast to the case of the bradyon (Fig. 11). However,
it
is not clear whether this modified view can provide the consistent covariant
theory of the \v{C}erenkov effect. Furthermore, we note that the effect depends
on the microscopic structure of the tachyon like as the self-energy problem
such
that we can imagine the complete understanding can be obtained only after the
quantization. So far our understanding for the \v{C}erenkov effect has been
restricted to the
usual coupling, ${\it i.e.,}$ minimal couplings. So it would be interesting to
study the \v{C}erenkov effect for the non-minimal coupling for the tachyon.
However, we note that the minimal coupling is remainly good coupling for the
case
of the ``charged'' luxon as in the case of the bradyon. In this
case, by considering the Lorentz invariant electric charge $q$ of the luxon
[31], the total action becomes
\begin{eqnarray}
{\cal J}[~\eta_{1},~ \eta_{2}~]&=&\int _{\eta_{1}}^{\eta_{2}}
\left[ -\frac{1}{2} \xi (\eta) ~l^{\mu} l_{\mu}  ~d \eta
-\frac{q}{c}  A_{\mu}(~x_{p}(\eta)~)~ dx^{\mu}_{p}
\right]
             \nonumber \\
&=&\int _{\eta_{1}}^{\eta_{2}}
\left[ -\frac{1}{2} \xi (\eta) ~\frac{d x^{\mu}}{d \eta}
\frac{d x_{\mu}}{d \eta} d \eta
-\frac{q}{c}  A_{\mu}(~x_{p}(\eta)~) ~dx^{\mu}_{p}
\right]
\end{eqnarray}
for the particle position $x^{\mu}_{p}(\eta)$, and the equation of motion
becomes
\begin{eqnarray}
\frac{d}{d \eta} \left[ \frac{d x_{\nu}}{d \eta} \xi (\eta) \right]
=\frac{q}{c} ~F_{\nu \mu} ~\frac{d x^{\mu}}{d \eta},
\end{eqnarray}
or with the physical time $t$
\begin{eqnarray}
\frac{d}{dt} P_{\nu}~=~q~ F_{\nu \mu} ~\frac{d x^{\mu}}{d t},
\end{eqnarray}
where $dx^{\mu}/dt=c(1,\hat{k})$. Note that this interaction does not induce
the \v{C}erenkov radiation in vacuum.

Now, for the quantization, it has been noted that a new method of
quantization is needed in the field quantization theory for the tachyon, but no
consistent method has been suggested so far [2,9]. Here we note that more
fundamentally the quantum theory of the tachyon, if it exists, is different
from
the usual quantum theory of the bradyon by considering the measurement of the
momentum and
coordinate of a tachyon.

To this end we first note that for the coordinate measurement at a time $t$
the uncertainty of the coordinate $\Delta {\bf x}$ for one object is related by
the uncertainty of the momentum $\Delta {\bf p}$ as
\begin{eqnarray}
\Delta {\bf x}(t)~ \geq ~\frac{\hbar}{ \Delta {\bf p}(t) }
\end{eqnarray}
according to the Heisenberg's uncertainty principle. Then, it is easy to see
that for the bradyon there is a lower bound for the magnitude of the
right-hand side of the relation (126) as $\hbar/(2m_{0}c)$ [32] since i) for
the
one
object case, $2m_{0}c$ is the maximum uncertainty of the momentum for it's
validity according to the usual particle and anti-particle theory
{\it indirectly} due to the existence of the maximum uncertainty for the
energy [33], and ii) for the
many object case, {\it i.e.}, when the many object and anti-object pairs are
produced from the high momentum transfer $\Delta {\bf p}$ in the measurement
process, $\Delta {\bf x}$ (uncertainty of coordinate for each object) behaves
as
$\sim~\Delta {\bf X}/N$ as $\Delta {\bf P}$ behaves as $\sim~ N m_{0}c$
($ \Delta {\bf X}$ and $\Delta {\bf P}$ represent the uncertainty of the
coordinate and momentum for the total system of the many objects respectively).
Hence, in the quantum theory it is, in principle, impossible to make an
arbitrarily accurate measurement of the coordinate of the bradyon. This
property is still satisfied even by the tachyon although the situation is
different. In this case there is also a lower bound {\it directly} due to the
existence of the maximum uncertainty of the momentum for maintaining the number
of the objects although there is no maximum uncertainty for the energy in that
case (Fig. 11, 12) both for the one object case and many object case similarly
to the analysis of the bradyon. Hence, as for the measurement of the coordinate
the tachyon and bradyon behave
similarly, {\it i.e.,} it is, in principle, impossible to make an arbitrarily
accurate measurement of the coordinate for the case of the both tachyon and
bradyon [34].

However, this is not case for the momentum measurement. To see this we note
that, following the argument of Landau and Peierls [32], for
the momentum measurement during a time interval $\Delta t$, the uncertainty of
the momentum $\Delta {\bf p}$ is related as
\begin{eqnarray}
\Delta {\bf p} \Delta t ~\geq ~\frac{\hbar}{({\bf u}-{\bf u}')}
\end{eqnarray}
by applying the Heisenberg's uncertainty principle, where ${\bf u}$ and
${\bf u}'$ are the velocities of the tachyon before and after the measurement.
Now it is easy to see that, for the bradyon, there is a lower bound for the
right-hand side of the relation (127) as $\hbar/c$, due to the velocity limit
of
the bradyon $c$, such that in the relativistic quantum theory it is, in
principle, impossible to make an arbitrarily accurate and ${\it rapid}$
measurement of the momentum. On the other hand, this is not true for the case
of the tachyon since in this case there is no upper velocity limit such that
any arbitrarily accurate and rapid measurement of the momentum.

Hence in the quantum theory, if it exists, the momentum
of the tachyon can act as the dynamical variable in contrast to the case of the
relativistic bradyon such that the physical significance of the momentum-space
wavefunction $\varphi ({\bf p})$ is as important as that of the
non-relativistic bradyon. However, the coordinate of the tachyon can not act as
the dynamical variable similar to the case of the relativistic bradyon such
that
the wavefunction $\psi ({\bf x},t)$ is not as important as that of the
non-relativistic case.  In this sense the quantum theory of the tachyon
resembles that of the non-relativistic quantum theory of the bradyon in one
aspect, {\it i.e.,} momentum measurement, but resembles that of the
relativistic quantum theory of the bradyon in another aspect, {\it i.e.,}
coordinate measurement. In other words, the quantum theory of the tachyon, if
it exists, would be a mixture of the relativistic and non-relativistic quantum
theory of the bradyon.
Furthermore, we are questioning whether, due to these peculiar properties, the
divergence problem of the usual quantum field theory of the bradyon does not
appear in the case of the tachyon since the problem is not independent on the
limit of the accuracy for the measurement of the coordinate and the momentum.

Finally, we hope that through further investigation the consistent dynamics
of the tachyon including the interaction  will be established.\\

\begin{center}
{\large \bf ACKNOWLEDGMENTS}
\end{center}
The present work was supported by the Basic Science Research Institute
program, Ministry of Education, Project No. BSRI-95-2414.

\newpage
\begin{center}
{\large \bf REFERENCES}
\end{center}
\begin{description}
\item{1.} It should be understood hereafter that all our analysis are made
in `` vacuum '' though not specified explicitly.

\item{2.} S. Tanaka, {\it Prog. Theor. Phys.} {\bf 24} (1960), 171;
O. M. P. Bilaniuk, V. K. Deshpande, and E. C. G. Sudarshan, {\it Am. J. Phys.}
{\bf 30} (1962), 718; G. Feinberg, {\it Phys. Rev. D} {\bf 159} (1967), 1089.

\item{3.} A. Einstein, {\it Ann. Phys.} {\bf 17} (1905), 891.

\item{4.} A. Einstein, {\it Jahrb. Radioaki.} {\bf 4} (1907), 411;
{\bf 5(E)} (1907), 98. The English version can be obtained from `` The
Collected Papers of Albert Einstein '' ( A. B. Beck and P. Havas, Eds.),
Vol. 2, p. 252, Princeton University Press, Princeton, New Jersey, 1989.

\item{5.} R. G. Cawley, {\it Ann. Phys.} {\bf 54} (1969), 122;
{\it Phys. Rev. D} {\bf 2}
(1970), 276; E. Recami, {\it Accad. Naz. Lincei Rendic. Sc.} {\bf 49} (1970),
77.

\item{6.} A. Einstein, `` Relativity '', 15th ed., p. 35, Crown Publishers,
Inc., New York, 1952; A. I. Miller, `` Albert Einstein's Special Theory of
Relativity '', p. 331, Addison- Wesley Publishing Company, Inc., Sydney, 1981.

\item{7.} This is because the ordinary detectors will get some imaginary
energy or momentum by detecting the tachyon since the process of the
measurement is essentially due to the exchange of the energy or momentum in a
detectable time interval.

\item{8.} O. M. Bilaniuk and E. C. G. Sudarshan, {\it Nature}, {\bf 223}
(1969),
386;
O. M. Bilaniuk et. al, {\it Phys. Today} {\bf 22} (No.12) (1969), 47; R. G.
Newton,
{\it Science} {\bf 167} (1970), 1569 and references therein.

\item{9.} M. E. Arons and E. C. G. Sudarshan, {\it Phys. Rev.} {\bf 173}
(1968),
1622;
J. Dhar and E. C. G. Sudarshan, {\it ibid.} {\bf 174} (1968), 1808;
G. Ecker, {\it Ann. Phys.} {\bf 58} (1970), 303; B. Schroer, {\it Phys. Rev. D}
{\bf 3}
(1971), 1764; K. Kamoi and S. Kamefuchi, {\it Prog. Theor. Phys.} {\bf 45}
(1971),
1646; C. Jue, {\it Phys. Rev. D} {\bf 8} (1973), 1757.

\item{10.} D. Korff and Z. Fried, {\it Nuov. Cim.} {\bf 52} (1967), 173; Y. P.
Terletskii, `` Paradoxes in The Theory of Relativity '', p. 88, Plenum Press,
New York,
1968.

\item{11.} O. M. Bilaniuk and E. C. G. Sudarshan, {\it Phys. Today} {\bf 22}
(No.5)
(1969), 43.

\item{12.} The reverse of this reasoning is also true, {\it i.e.}, if Eq. (3)
(energy) is defined to the totally conserved quantity, mass $m({\bf u})$ is
found as Eq. (2), and Eq. (1) (momentum) is proved to be totally conserved.
See, for example, R. Penrose and W. Rindler, {\it Am. J. Phys.} {\bf 33}
(1965),
55;
J. Ehlers, W. Rindler, and R. Penrose, {\it ibid}. {\bf 33} (1965), 995.

\item{13.} P. G. Bergmann, `` An Introduction to the Theory of Relativity '',
p. 87, Prentice-Hall International, Inc., London, 1942; C. M{\o}ller, `` The
Theory of Relativity '', p. 67, Oxford University Press, Oxford, London, 1952);
M. Born, `` Einstein's Theory of Relativity '', p. 267 , Dover Publications,
Inc., New York, 1962.

\item{14.} G. N. Lewis and R. C. Tolman, {\it Phil. Mag.} {\bf 18} (1909), 510;
R. C. Tolman, {\it ibid.} {\bf 23} (1912), 375; `` Relativity, Thermodynamics
and Cosmology '', p. 42, Oxford University Press, Oxford, England, 1934;
A. Einstein, {\it Amer. Math. Soc. Bull.} {\bf April} (1935), 223;
W. H. McCrea, `` Relativity Physics '', p. 19, Methuen and Company, Ltd.,
London, 1935;
D. Bohm, `` The Special Theory of Relativity '', p. 81, W. A. Benjamin, Inc.,
New York, 1965; R. K. Pathria, `` The Theory of Relativity '', 2nd ed., p. 77,
Pergamon Press Ltd., Headington Hill Hall, Oxford, 1974;
D. S. Mann and P. K. Mukherjee, `` Relativity, Mechanics and Statistical
Physics '', p. 50, John Wiley and Sons, Inc., New York. The full
review of the various other formulations as well as afore-mentioned the two can
be obtained from H. Arzeli{\` e}s, `` Relativistic Point Dynamics '', p. 19,
Pergamon Press, Oxford, 1971.

\item{15.} H. Arzeli{\` e}s [14].

\item{16.} L. Parker, {\it Phys. Rev.}  {\bf 188} (1969), 2287;
E. Recami and R. Mignani, {\it Lett. Nuov. Cim.} {\bf 4} (1972), 144;
A. F. Antippa, {\it Nuov. Cim. A} {\bf 10} (1972), 389;
R. Goldoni, {\it ibid.} {\bf 13} (1973), 501; 527;
H. C. Corben, {\it ibid.} {\bf 29} (1975), 415.

\item{17.} V. B. Berestetski\v{i}, E. M. Lifshitz, and L. P. Pitaerski\v{i},
`` Relativistic Quantum Theory '', Pergamon Press, New York, 1971.

\item{18.}
Although several authors in Ref.[10] pointed out the incompleteness of the
usual
formulation of BDSF with regard to the problem of the
conservation law of momentum and energy, they did not provide the complete
solution to this problem in contrast to their claims.

\item{19.} L. D. Landau and E. M. Lifshitz, `` The Classical Theory of
Fields '', 4th English ed., Pergamon Press, Oxford, 1975; S. Weinberg,
`` Gravitation and Cosmology '', John Wiley \& Sons, New York, 1972;
F. Rohrlich, `` Classical Charged Particles '', Addison-Wesley Publishing
Company, Inc., Massachusetts, 1965.

\item{20.} Although the pseudo-tensor $T^{\mu \nu ...\omega}$ is
defined usually as
$T^{\mu \nu ...\omega}=\alpha {\Lambda^{\mu}}_{\mu'}{\Lambda^{\nu}}_{\nu'} ...
{\Lambda^{\omega}}_{\omega'}T^{\mu' \nu' ...\omega'}$,
$\alpha={\mbox det}({\Lambda^{\mu}}_{\nu})= \pm 1 $
with the orthogonal transformation matrix ${\Lambda^{\mu}}_{\nu}$, we introduce
here more general definition of it having the arbitrary sign-factor $\alpha$
which is
not necessarily the ${\mbox det}({\Lambda^{\mu}}_{\nu})$. In our case $\alpha$
is the anomalous sign-factor `` $\mbox{sign} \{ 1 -{\bf u}\cdot {\bf v}/{c^{2}}
 \}$ ''.

\item{21.} E. C. G. St\"{u}eckelberg, {\it Helv. Phys. Acta.} {\bf 14} (1941),
588;
{\it ibid.} {\bf 15} (1942), 23( We thank Jyung-Youn Choi for English
translation from French); R. P. Feynman, {\it Phys. Rev.} {\bf 74} (1948), 939;
{\it ibid.} {\bf 76} (1949), 749.

\item{22.} For the free object, the imaginary-valued action may be allowed
with slight modification of the action principle without changing the physical
contents. However, only the real-valued action is meaningful if we consider
the physically meaningful interactions between this system and the system of
the bradyon such that the imaginary-valued momentum and energy does not appear
in the theory if the action of the bradyon is defined to be real-valued as
usual.

\item{23.} Although it seems that the derivation of Eq. (76) has some
ambiguities since we must use the relation $(d \lambda/dt) dt =d \lambda$,
which is ambiguous for $dt=0$ case. But this is actually not the case because
the last formula exactly reproduce the first one even in this case.

\item{24.} Of course, the explicit pattern of
$ m_{0} \rightarrow 0$ as $u \rightarrow c$ can not be given according to the
results of the previous two sections because $m_{0}$ is defined to be Lorentz
constant without the functional dependence of any other variable, like as it's
velocity.

\item{25.} Of course, it is possible to prove that this formula by using the
Maxwell equation is satisfied for the chunks of the electromagnetic radiation.
See, for example, C. M{\o}ller [13] and Ref. [19].

\item{26.} Here we are not considering the internal spin quantity for the
luxon.

\item{27.} M. B. Green, J. H. Schwartz and E. Witten, `` Superstring Theory ''
Vol. 1, Cambridge University Press, Cambridge, 1987; A. M. Polyakov,
`` Gauge Fields and Strings '', Harwood Academic Publishers, England, 1987.

\item{28.} In general, it is expected that the action produces trivial or
non-dynamical information about the interesting system if we implement the
equations of motion with respect to variation of some dynamical variables back
into the action. This is essentially due to the fact that the equations of
motion are nothing but the conditions for the stable point of the action, which
is non-dynamical.

\item{29.} A. Sommerfeld, {\it Proc. Roy. Acad. Amsterdam}, {\bf 7} (1904),
346;
P. A. \v{C}erenkov, {\it Phys. Rev.} {\bf 52} (1937), 378 and references
therein;
I. M. Frank and I. E. Tamm, {\it C. R. Ac. Sci. U.S.S.R.} {\bf 14} (1937), 109;
E. C. G. Sudarshan [2].

\item{30.} There have been several controversies about the \v{C}erenkov effect
in the vacuum. E. C. G. Sudarshan, {\it Ark. Phys.} {\bf 39} (1970), 40;
T. Alv\"{a}ger and M. N. Kreisler, {\it Phys. Rev.} {\bf 171} (1968), 1357;
F. Jones, {\it Phys. Rev. D} {\bf 6} (1972), 2727;
H. K. Wimmel, {\it Lett. Nuov. Cim.} {\bf 1} (1971), 645;
{\it Nature Phys. Sci.} {\bf 236} (1972), 79;
R. Mignani and E. Recami, {\it Lett. Nuov. Cim.}  {\bf 7} (1973), 388.

\item{31.} Although, as far as we know, there is no electrically charged luxon,
we are considering this object since there is no reason to reject this
possibility.

\item{32.} L. D. Landau and R. Peierls, {\it Z. Phys.} {\bf 69} (1931), 56; L.
D.
Landau and E. M. Lifshitz, `` Quantum Mechanics '', \S 44, Addison-Wesley,
Reding, Massachusetts, 1977.

\item{33.} By using the energy-momentum relation for the bradyon
$E=\sqrt{p^{2}c^{2}+m_{0}^{2}c^{4}}$, the inequality
$1/\Delta p \geq c/\Delta E >1/(2 m_{0} c)$ is resulted due to the existence of
the maximum
uncertainty of the energy for the validity of one object case.

\item{34.} This property has been already known in different context,
A. Peres, {\it Lett. Nuov. Cim.} {\bf 1} (1969), 837.

\end{description}
\newpage
\begin{center}
{\large \bf FIGURES CAPTIONS}
\end{center}

Fig. 1 : For the case of the bradyon, all the directions of the velocity of the
moving object relative to the rest object in $S$ are equivalent due to the
isotropy of the space. So, if there is the mass relation between the
objects moving relative to the rest one for the same kind of objects,
the transformation function should not depend on the direction of the velocity
of the moving object. \\

Fig. 2 : For the case of the tachyon, all the directions of the velocity of
the general tachyons relative to the tachyon with $u_{a}^{2}=2 c^{2}$ in $S$
are
not equivalent due to the asymmetry of the situation. So, if there is any
relation between the masses of the general tachyons moving relative to the
tachyons with
$u_{a}^{2}=2 c^{2}$ for the same kinds, the function can be depend on the
velocity of the tachyon relative to
the tachyon with $u_{a}^{2}=2 c^{2}$ in general. \\

Fig. 3 : If the mass of the tachyon depends on the direction of the velocity
for a given speed, it should be also changed for the different orientations of
the coordinates. But this is not allowed in order that the total energy and
momentum of Eqs. (4) and (5) are conserved in any orientation of coordinates
because the bradyon's mass is independent on the direction of the velocity. \\

Fig. 4 : In order that the momentum is also conserved in the system
$S'$, $\sum_{a} \epsilon_{a}$ should be conserved, {\it i.e.},
$\sum^{M}_{a_{i}=1} \epsilon_{a_{i}}=\sum^{N}_{a_{f}=1} \epsilon_{a_{f}}$
according to Eq. (25) or (26), as well as the energy and momentum
conservation. \\

Fig. 5 : For the bradyonic world-line, the world-velocity $u^{\mu}$ is
time-like, {\it i.e.}, $u^{\mu} u_{\mu}$ = $c^{2} ~>~0 $, while the
world-acceleration $a^{\mu}$ is space-like, ${\it i.e.}$,
$a^{\mu}a_{\mu}=-a^{2} c^{2} \{1-(u^{2}/c^{2})
\mbox{sin}^{2} \phi\} /\{c^{2}(1-u^{2}/c^{2})^{3} \} ~<~0 $ for any angle
$\phi$ between ${\bf a}$ and ${\bf u}$. \\

Fig. 6 : For the tachyonic world-line, the world-velocity $t^{\mu}$ is
space-like, {\it i.e.}, $t^{\mu}t_{\mu}=-c^{2}~<~0$, while the
world-acceleration
$a^{\mu}_{t}$ is space-like or time-like or light-like, {\it i.e.}, the sign of
$a^{\mu}_{t}{a_{\mu}}_{t}=-a^{2} c^{2} \{1-(u^{2}/c^{2})
\mbox{sin}^{2} \phi\} /\{c^{2}(1-u^{2}/c^{2})^{3} \} $ depends on the angle
$\phi$ for a given velocity ${\bf u}$. \\

Fig. 7 : Virtual variation of the bradyonic trajectory in the space-time
diagram with the conditions $\delta {\bf x}(t_{1})=\delta {\bf x}(t_{2})=0$.
The
virtual
variation $\delta x$ defines the variation with fixed time $t$ such that
$\delta (d{\bf x}/dt)=d(\delta {\bf x})/dt$. Furthermore, the variations are
restricted to
the time-like cone. \\

Fig. 8 : Usual variation when applied to the tachyonic trajectory. However,
this
method with fixed time is doubtful for the infinite velocity tachyon, where the
time interval always vanish such that there are some ambiguities in deriving
the equations of motion. Furthermore, since any tachyon can be made to have the
infinite velocity by choosing an appropriate system of coordinates, this
problem
can not be avoided for any tachyon. \\

Fig. 9 : Variation in ($\lambda, x$) space for the tachyonic trajectory.
In this space there are no ambiguities for any tachyon:
the ambiguity at $u=\infty$ is transferred to the ambiguity at $u=c$,
{\it i.e.,} the case of the luxon such that there is no problem for the
tachyon. The allowed trajectories are restricted to the space-like cone except
$\lambda =0$ plane. \\

Fig. 10 : A possible function dependence of $t$ and $\lambda$ for the
infinite-velocity
tachyon according to the relation (43). In this case the physical time $t$ is
(constant)
function of $\lambda$, but $\lambda$ is not a function of $t$ in mathematical
sense. \\

Fig. 11 : There is no energy gap in the energy spectrum of the tachyon ((b)) in
contrast to the energy gap $2|m_{0}|c^{2}$ for the bradyon ((a)). \\

Fig. 12 : There exists the threshold momentum $\kappa c$ for the creation of
the
tachyon ((b)) in
contrast to the case of the bradyon ((a)).

\end{document}